\documentclass[11pt,leqno]{article}

\usepackage{mathptmx}

\usepackage{amssymb,amsmath,amsthm}
\usepackage{comment}
\usepackage[latin1]{inputenc}    
\usepackage{pifont}    

\makeatletter%

\usepackage{mathptmx}
\newcommand{\doublespacing}{\let\CS=
\@currsize\renewcommand{\baselinestretch}{1.75}\tiny\CS}
\newcommand{\extradoublespacing}{\let\CS=
\@currsize\renewcommand{\baselinestretch}{1.9}\tiny\CS}
\newcommand{\draftspacing}{\let\CS=
\@currsize\renewcommand{\baselinestretch}{2.0}\tiny\CS}
\newcommand{\hugedraftspacing}{\let\CS=
\@currsize\renewcommand{\baselinestretch}{2.4}\tiny\CS}
\makeatother%

\newcommand{\OMIT}[1]{} %
\newcommand{\jfootnote}[1]{} %
\newcommand{\gfootnote}[1]{} %

\newtheorem{rules}{Rule}

\newcommand\qedblob{\ding{113}}
\def\literalqed{{\ \nolinebreak\hfill\mbox{\qedblob\quad}}}

\def\qed{\literalqed}
\newenvironment{proofs}{\noindent{\bf Proof.}\hspace*{1em}}{\literalqed\bigskip}
\newcommand{\sproofsketchof}[1]{\noindent{\bf Proof Sketch of {#1}.}\hspace*{1em}}
\newcommand{\eproofof}[1]{\hfill \mbox{\qed\quad$_{\mbox{\small {#1}}}$}\medskip}

\newcommand{\seq}{\subseteq}

\newcommand{\scoresub}[2]{\mathit{score}_{#1}(#2)}

\newcommand{\diff}{\mathit{diff}}

\newcommand{\np}{\mbox{\rm NP}}

\newcommand{\condition}{\,|\:}

\newenvironment{desctight}
  {\begin{list}{}{\setlength\labelwidth{0pt}%
        \setlength{\itemsep}{0.5pt}%
        \setlength{\parsep}{0pt}%
        \setlength\itemindent{-\leftmargin}%
        }}
    {\end{list}}

  \newtheorem{theorem}{Theorem}[section]

  \newtheorem{lemma}[theorem]{Lemma}

  \newtheorem{proposition}[theorem]{Proposition}
  \newtheorem{definition}[theorem]{Definition}
  
  \newtheorem{construction}[theorem]{Construction}
\usepackage{ulem}
\begin{document}

\title{
Sincere-Strategy Preference-Based Approval Voting Fully Resists
  Constructive Control and Broadly Resists Destructive
  Control\thanks{Supported in part by the DFG under grants 
RO~\mbox{1202/12-1} (within the European Science
Foundation's EUROCORES program LogICCC: 
``Computational Foundations of Social Choice'') and RO~\mbox{1202/11-1}
  and by the Alexander von Humboldt Foundation's
  TransCoop program.  
Preliminary versions of this paper have been presented at
  the 33rd International Symposium on Mathematical Foundations of
  Computer Science (MFCS-08)~\cite{erd-now-rot:c:sp-av}
  and at the 2nd International Workshop on Computational Social
  Choice (COMSOC-08). Work done in part while the first author
  was visiting Universit\"{a}t Trier and while the third author
  was visiting the University of Rochester.}}

\author{G\'{a}bor Erd\'{e}lyi\thanks{URL: 
\mbox{\tt{}ccc.cs.uni-duesseldorf.de/\mbox{\tiny$\sim\,$}erdelyi.}
}
\ and \
Markus Nowak
\ and \
J\"{o}rg Rothe\thanks{URL: 
\mbox{\tt{}ccc.cs.uni-duesseldorf.de/\mbox{\tiny$\sim\,$}rothe.}
} \\
Institut f\"{u}r Informatik \\
Heinrich-Heine-Universit\"{a}t D\"{u}sseldorf \\
40225 D\"{u}sseldorf \\
Germany
}

\date{June 12, 2008}

\maketitle

\begin{abstract}
  We study sincere-strategy preference-based approval voting (SP-AV),
  a system proposed by Brams and
  Sanver~\cite{bra-san:j:critical-strategies-under-approval} and
  here adjusted so as to \emph{coerce} admissibility of the votes
  (rather than excluding inadmissible votes \emph{a priori}),
  with respect to
  procedural control.  In such control scenarios, an external agent
  seeks to change the outcome of an election via actions such as
  adding/deleting/partitioning either candidates or voters.  SP-AV
  combines the voters' preference rankings with their approvals of
  candidates, where in elections with at least two candidates the voters'
  approval strategies are adjusted---if needed---to approve of their
  most-preferred candidate and to disapprove of their least-preferred
  candidate.  
  This rule coerces admissibility of the votes even in the presence of control
  actions, and hybridizes, in effect, approval with pluralitiy voting.

  We prove that this system is computationally resistant
  (i.e., the corresponding control problems are $\np$-hard) to $19$
  out of $22$ types of constructive and destructive control.  Thus,
  SP-AV has more resistances
  to control
  than is currently known for any other natural
  voting system with a polynomial-time winner problem.  In particular,
  SP-AV is (after Copeland
  voting, see Faliszewski et
  al.~\cite{fal-hem-hem-rot:j-To-Appear-With-TR-Ptr:llull-copeland-full-techreport,fal-hem-hem-rot:c:copeland-fully-resists-constructive-control})
  the second natural voting system with an easy winner-determination
  procedure that is known to have full resistance to constructive
  control, and unlike Copeland voting it in addition displays broad
  resistance to destructive control.
\end{abstract}
\maketitle                   %

\section{Introduction}
\label{sec:intro}

Voting provides a particularly useful method for preference
aggregation and collective decision-making.  While voting systems were
originally used in political science, economics, and operations
research, they are now also of central importance in various areas of
computer science, such as artificial intelligence (in particular,
within multiagent systems).  In automated, large-scale computer
settings, voting systems have been applied, e.g., for
planning~\cite{eph-ros:c:multiagent-planning} and similarity
search~\cite{fag-kum-siv:c:similarity-search}, and have also been used
in the design of recommender
systems~\cite{gho-mun-her-sen:c:voting-for-movies} and ranking
algorithms~\cite{dwo-kum-nao-siv:c:rank-aggregation} (where they help
to lessen the spam in meta-search web-page rankings).  For such
applications, it is crucial to explore the computational properties of
voting systems and, in particular, to study the complexity of problems
related to voting (see, e.g., the survey by Faliszewski et
al.~\cite{fal-hem-hem-rot:b:richer}).

The study of voting systems from a complexity-theoretic perspective
was initiated by Bartholdi, Tovey, and Trick's series of seminal
papers about the complexity of winner
determination~\cite{bar-tov-tri:j:who-won},
manipulation~\cite{bar-tov-tri:j:manipulating},
and procedural control~\cite{bar-tov-tri:j:control} in elections.
This paper contributes to the study of electoral control, where an
external agent---traditionally called \emph{the chair}---seeks to
influence the outcome of an election via procedural changes to the
election's structure, namely via adding/deleting/partitioning either
candidates or voters (see Section~\ref{sec:prelims:control-problems}
for the formal definitions of our control problems).  We consider both
\emph{constructive} control (introduced by Bartholdi, Tovey, and
Trick~\cite{bar-tov-tri:j:control}), where the chair's goal is to make
a given candidate the unique winner, and \emph{destructive} control
(introduced by Hemaspaandra, Hemaspaandra, and
Rothe~\cite{hem-hem-rot:j:destructive-control}), where the chair's
goal is to prevent a given candidate from being a unique winner.

We investigate the same twenty types of constructive and destructive
control that were studied for approval
voting~\cite{hem-hem-rot:j:destructive-control} and two additional
control types introduced by Faliszewski et
al.~\cite{fal-hem-hem-rot:c:llull} (see also
\cite{fal-hem-hem-rot:j-To-Appear-With-TR-Ptr:llull-copeland-full-techreport}),
and we do so for a variant of a
voting system that was proposed by Brams and
Sanver~\cite{bra-san:j:critical-strategies-under-approval} as a
combination of preference-based and approval voting.  Approval voting
was introduced by Brams and Fishburn~\cite{bra-fis:j:approval-voting}
as follows: Every voter
either approves or disapproves of each candidate, and every candidate
with the largest number of approvals is a winner.  One of the simplest
preference-based voting systems is plurality: All voters report their
preference rankings of the candidates, and the winners are the
candidates that are ranked first-place by the largest number of
voters.  The purpose of this paper is to show that Brams and Sanver's
combined system (adapted here so as to keep its useful features even
in the presence of control actions) combines the strengths, in terms
of computational resistance to control, of plurality and approval
voting.

Some voting systems are \emph{immune} to certain types of control in
the sense that it is never possible for the chair to reach his or her
goal via the corresponding control action.
Immunity to any
type of control
unconditionally shields the
voting system against this particular control type.
However,
like most voting systems approval voting is \emph{susceptible} (i.e.,
not immune) to many types of control, and plurality voting is
susceptible to all types of control.\footnote{A related line of
research has shown that, in principle, all natural voting systems can
be manipulated by strategic voters.  Most notable among such results
is the classical work of
Gibbard~\cite{gib:j:manipulation-voting-schemes} and
Satterthwaite~\cite{sat:j:strategy-proofness}.  The study of
strategy-proofness is still an extremely active and interesting area
in social choice theory (see, e.g., Duggan and
Schwartz~\cite{dug-sch:j:polsci:gibbard}) and in artificial
intelligence (see, e.g., Everaere et
al.~\cite{eve-kon-mar:j:strategy-proofness-landscape-of-merging}).}
However, and this was Bartholdi,
Tovey, and Trick's brilliant insight~\cite{bar-tov-tri:j:control},
even for systems susceptible to control, the chair's task of
controlling a given election may be too hard computationally (namely,
$\np$-hard) for him or her to succeed.  The voting system is then said
to be \emph{resistant} to this control type.  If a voting system is
susceptible to some type of control, but the chair's task can be
solved in polynomial time, the system is said to be \emph{vulnerable}
to this control type.

The quest for a natural voting system with an easy
winner-determination procedure that is universally resistant to
control has lasted for more than 15 years now.  Among the voting systems
that have been studied with respect to control are plurality,
Condorcet, approval, cumulative, Llull, and (variants of) Copeland
voting~\cite{bar-tov-tri:j:control,hem-hem-rot:j:destructive-control,hem-hem-rot:c:hybrid,mei-pro-ros-zoh:j:multiwinner,fal-hem-hem-rot:c:llull,fal-hem-hem-rot:c:copeland-fully-resists-constructive-control,bet-uhl:c:parameterized-complecity-candidate-control,fal-hem-hem-rot:j-To-Appear-With-TR-Ptr:llull-copeland-full-techreport}.
Among these systems, plurality and Copeland voting (denoted
Copeland$^{0.5}$
in~\cite{fal-hem-hem-rot:j-To-Appear-With-TR-Ptr:llull-copeland-full-techreport,fal-hem-hem-rot:c:copeland-fully-resists-constructive-control})
display the broadest resistance to control, yet even they are not
universally control-resistant.  The only system currently known to be
fully resistant---to the $20$ types of constructive and destructive
control studied
in~\cite{hem-hem-rot:j:destructive-control,hem-hem-rot:c:hybrid}---is
a highly artificial system constructed
via hybridization~\cite{hem-hem-rot:c:hybrid}.
(We
mention that this system was not designed for direct, real-world
use as a ``natural'' system but rather was intended to rule out the
existence of a certain impossibility theorem~\cite{hem-hem-rot:c:hybrid}.)

While approval voting nicely distinguishes between each voter's
acceptable and inacceptable candidates, it
ignores the preference rankings the voters may have about their
approved (or disapproved) candidates.  This shortcoming motivated
Brams and Sanver~\cite{bra-san:j:critical-strategies-under-approval}
to introduce a voting system that
combines approval and preference-based voting, and they defined the
related notions of sincere and admissible approval strategies, which
are quite natural requirements.
We adapt their sincere-strategy preference-based approval voting
system in a natural way such that, for elections with at least two
candidates, admissibility of approval strategies (see
Definition~\ref{def:sv-ap}) can be ensured even in the presence of
control actions such as deleting candidates and partitioning
candidates or voters.\footnote{Note that in control by partition of
voters (see Section~\ref{sec:prelims:control-problems}) the run-off
may have a reduced number of candidates.}
The purpose of this paper is to study if, and to what extent, this
system
inherits
the control resistances of plurality (which is perhaps the simplest
preference-based system) and approval voting.  Denoting this system by
SP-AV, we show that SP-AV does combine all the resistances of
plurality and approval voting.

\begin{table}[t+]
\centering
\begin{tabular}{|l||c|c|c|c|c|c|}
\hline
Number of & Condorcet & Approval & Llull & Copeland & Plurality  & SP-AV \\
\hline\hline
resistances     &  $3$ &  $4$ & $14$ & $15$ & $16$ &  $19$ \\ \hline
immunities      &  $4$ &  $9$ &  $0$ &  $0$ &  $0$ &   $0$ \\ \hline
vulnerabilities &  $7$ &  $9$ &  $8$ &  $7$ &  $6$ &   $3$ \\ \hline\hline
References      &
\cite{bar-tov-tri:j:control,hem-hem-rot:j:destructive-control} &
\cite{bar-tov-tri:j:control,hem-hem-rot:j:destructive-control} &
\cite{fal-hem-hem-rot:j-To-Appear-With-TR-Ptr:llull-copeland-full-techreport,fal-hem-hem-rot:c:llull,fal-hem-hem-rot:c:copeland-fully-resists-constructive-control} &
\cite{fal-hem-hem-rot:j-To-Appear-With-TR-Ptr:llull-copeland-full-techreport,fal-hem-hem-rot:c:llull,fal-hem-hem-rot:c:copeland-fully-resists-constructive-control} &
\cite{bar-tov-tri:j:control,hem-hem-rot:j:destructive-control,fal-hem-hem-rot:j-To-Appear-With-TR-Ptr:llull-copeland-full-techreport,fal-hem-hem-rot:c:llull} &
Theorem~\ref{thm:summary-of-results} \\
\hline
\end{tabular}
\caption{\label{tab:number-of-resistances}
Number of resistances, immunities, and vulnerabilities to
our $22$ control types.  (Regarding the ``Condorcet'' column, see
Footnote~\ref{foo:condorcet}.)}
\end{table}

More specifically, we prove that sincere-strategy preference-based
approval voting is resistant to $19$ and vulnerable to only three of
the $22$ types of control considered here.
For
comparison, Table~\ref{tab:number-of-resistances} shows the number of
resistances, immunities, and vulnerabilities to our $22$ control types
that are known for each of Condorcet,\footnote{\label{foo:condorcet}
Note that
Table~\ref{tab:number-of-resistances} lists only $14$ instead of $22$
types of control for Condorcet.  The reason is that, as
in~\cite{hem-hem-rot:j:destructive-control}, we consider two types of
control by partition of candidates (namely, with and without run-off)
and one type of control by partition of voters, and for each partition case
we use the rules TE (``ties eliminate'') and TP (``ties promote'')
for handling ties that may occur in the corresponding subelections 
(see Section~\ref{sec:prelims:control-problems}).  
However, since Condorcet winners are
always unique when they exist, the distinction between TE and TP is
not made for the partition cases within Condorcet voting.
Note further that the two additional control types 
in Section~\ref{sec:control-by-adding-candidates}
(namely, constructive and destructive control by adding a
limited number of candidates~\cite{fal-hem-hem-rot:j-To-Appear-With-TR-Ptr:llull-copeland-full-techreport,fal-hem-hem-rot:c:llull}) have not
been considered for Condorcet 
voting~\cite{bar-tov-tri:j:control,hem-hem-rot:j:destructive-control}.}
approval, Llull, plurality,\footnote{Regarding the references given
in Table~\ref{tab:number-of-resistances} for plurality, Faliszewski et
al.~\cite{fal-hem-hem-rot:j-To-Appear-With-TR-Ptr:llull-copeland-full-techreport,fal-hem-hem-rot:c:llull} note that plurality is resistant to constructive
and destructive control by adding a limited number of candidates (see
Section~\ref{sec:prelims:control-problems} for the definition of this
problem).  Hemaspaandra et
al.~\cite{hem-hem-rot:j:destructive-control} obtained all other results
for destructive control within plurality, and for the constructive
partitioning control cases in models TE and {TP}.  The remaining results
for plurality are due to Bartholdi et al.~\cite{bar-tov-tri:j:control}.
\label{foo:plurality}}
and Copeland voting (see
\cite{bar-tov-tri:j:control,hem-hem-rot:j:destructive-control,fal-hem-hem-rot:j-To-Appear-With-TR-Ptr:llull-copeland-full-techreport,fal-hem-hem-rot:c:llull,fal-hem-hem-rot:c:copeland-fully-resists-constructive-control}),
and for SP-AV (see Theorem~\ref{thm:summary-of-results} and
Table~\ref{tab:summary-of-results} in
Section~\ref{sec:results}).

This paper is organized as follows.  In Section~\ref{sec:prelims}, we
define and discuss
sincere-strategy preference-based approval voting, the types of
control studied in this paper, and the notions of immunity,
susceptibility, vulnerability, and resistance.
In Section~\ref{sec:results}, we prove our results on {SP-AV}.
Finally, in Section~\ref{sec:conclusions} we give our conclusions and
state some open problems.
\OMIT{
Finally, in
Section~\ref{sec:results:k-approval}, we study the complexity of
control for $k$-approval and sincere-strategy preference-based
$k$-approval voting ($k$-AV and SP-$k$-AV, for short).
} %

\section{Preliminaries}
\label{sec:prelims}

\subsection{Preference-Based Approval Voting}
\label{sec:prelims:variants-of-approval-voting}

An election $E = (C,V)$ is specified by a finite set $C$ of candidates
and a finite collection $V$ of voters who express their preferences
over the candidates in~$C$, where distinct voters may, of course, have
the same preferences.  How the voter preferences are represented
depends on the voting system used.  In approval voting (AV, for
short), every voter draws a line between his or her acceptable and
inacceptable candidates (by specifying a $0$-$1$ approval vector,
where $0$ represents disapproval and $1$ represents approval), yet
does not rank them.  In contrast, many other important voting systems
(e.g., Condorcet voting, Copeland voting, all scoring protocols,
including plurality, Borda count, veto, etc.) are based on voter
preferences that are specified as tie-free linear orderings of the
candidates.
As is most common in the literature, votes will here be
represented nonsuccinctly: one ballot per voter.  Note that some
papers (e.g.,
\cite{fal-hem-hem:c:bribery,fal-hem-hem-rot:j-To-Appear-With-TR-Ptr:llull-copeland-full-techreport,fal-hem-hem-rot:c:llull,fal-hem-hem-rot:c:copeland-fully-resists-constructive-control})
also consider succinct input representations for elections where
multiplicities of votes are given in binary.

Brams and Sanver~\cite{bra-san:j:critical-strategies-under-approval}
introduced a voting system that combines approval and preference-based
voting.  To distinguish this system from other systems that these
authors introduced with the same purpose of combining approval and
preference-based voting~\cite{bra-san:j:preference-approval-voting},
we call the variant considered here (including the
assumption of sincerity as explained below and
including Rule~\ref{rul:preference-rewrite-rule} below, which will coerce
admissibility)
\emph{sincere-strategy preference-based
  approval voting} (SP-AV, for short).

\begin{definition}[Brams and 
Sanver~\cite{bra-san:j:critical-strategies-under-approval}]
\label{def:sv-ap}
Let $(C,V)$ be an election, where the voters both indicate
approvals/\allowbreak{}disapprovals 
of the candidates and provide a tie-free linear
ordering of all candidates.  For each voter $v \in V$, an \emph{AV
  strategy of $v$} is a subset $S_v \seq C$ such that $v$ approves of
all candidates in $S_v$ and disapproves of all candidates in $C-S_v$.
The list of AV strategies for all voters in $V$ is called an \emph{AV
  strategy profile for $(C,V)$}.  (We sometimes also speak of
\emph{$V$'s AV strategy profile for $C$}.)  For each $c \in C$, let
$\scoresub{(C,V)}{c} = \|\{v \in V \condition c \in S_v\}\|$ denote
the number of $c$'s approvals. Every candidate $c$ with the largest
$\scoresub{(C,V)}{c}$ is a winner of election $(C,V)$.

An AV strategy $S_v$ of a voter $v \in V$ is said to be
\emph{admissible} if $S_v$ contains $v$'s most-preferred candidate and
does not contain $v$'s least-preferred candidate.\footnote{Brams and
Sanver~\cite{bra-san:j:critical-strategies-under-approval}
define an AV strategy to be admissible if it is not dominated
in a game-theoretic sense~\cite{bra-fis:j:approval-voting},
and note that ``admissible strategies under AV
involve always voting for a most-preferred candidate and
never voting for a least-preferred candidate.''  Since we do not focus
on the game-theoretic aspects of AV strategies, we define admissibility
as in Definition~\ref{def:sv-ap}.}
$S_v$ is said to be
\emph{sincere} if for each $c \in C$, if $v$ approves of $c$ then $v$
also approves of each candidate ranked higher than $c$ (i.e., there
are no gaps allowed in sincere approval strategies).  An AV strategy
profile for $(C,V)$ is \emph{admissible} (respectively,
\emph{sincere}) if the AV strategies of all voters in $V$ are
admissible (respectively, sincere).
\end{definition}

Admissibility and sincerity are quite natural requirements.  In
particular, requiring the voters to be sincere ensures that their
preference rankings and their approvals/disapprovals are not
contradictory.
Note that sincere strategies for at least two candidates are always
admissible if voters are neither allowed to approve of everybody nor
to disapprove of everybody (i.e., if we require voters $v$ to have
only AV strategies $S_v$ with $\emptyset \neq S_v \neq C$), an
assumption adopted by Brams and
Sanver~\cite{bra-san:j:critical-strategies-under-approval}.\footnote{%
\label{foo:admissible}
Brams and Sanver
actually preclude only the case $S_v = C$ for sincere 
voters~$v$ by stating
that ``sincere strategies are always admissible if we exclude `vote
for everybody'\,''~\cite{bra-san:j:critical-strategies-under-approval}.
However, an AV
strategy that disapproves of all candidates obviously is sincere, yet
not admissible according to Definition~\ref{def:sv-ap},
which is why we also exclude the case of $S_v = \emptyset$.}
Henceforth, we will
assume that only sincere AV strategy
profiles are considered, which---assuming
that the trivial cases $S_v = \emptyset$ and
$S_v = C$ are excluded---necessarily are admissible whenever there
are at least two candidates.\footnote{Note that an AV strategy is
never admissible for less than two candidates.}
A vote with an insincere strategy will be considered void.

The following notation was used by Brams and Sanver for a different
election system~\cite{bra-san:j:preference-approval-voting}, but is
useful for SP-AV as well:
Preferences are represented by a left-to-right ranking (separated by a
space) of the candidates (e.g., $\begin{array}{@{}c@{\ \ }c@{\ \
    }c@{}} a & b & c\end{array}$), with the leftmost candidate being
the most-preferred one, and approval strategies are denoted by
inserting a straight line into such a ranking, where all candidates
left of this line are approved of and all candidates right of this line
are disapproved of (e.g., ``$\begin{array}{@{}c@{\ \ }c@{\ \ }c@{\ \
    }c@{}} a & | & b & c\end{array}$'' means that $a$ is approved of,
while both $b$ and $c$ are disapproved of by this voter).
In our constructions, we
sometimes also insert a subset $B \seq C$ into such approval rankings,
where we assume some arbitrary, fixed order of the candidates in $B$
(e.g., ``$\begin{array}{@{}c@{\ \ }c@{\ \ }c@{\ \ }c@{}} a & | & B &
  c\end{array}$'' means that $a$ is approved of, while all $b \in B$ and
$c$ are disapproved of by this voter).

\OMIT{
\jfootnote{Definition $k$-AV und SP-$k$-AV to be OMITted in MFCS
version, but should be in journal version.}
\begin{definition}
  Let $k \geq 1$ be a fixed integer. In \emph{$k$-approval voting}
  ($k$-AV, for short), every voter approves of exactly $k$ of the $m
  \geq k$ candidates,\footnote{When there are less than $k$
  candidates, $k$-approval voting is not applicable.}  and all
  candidates with the largest number of approvals win.
  \emph{Sincere-strategy preference-based $k$-approval voting}
  (SP-$k$-AV, for short) in addition requires the voters to either
  approve or disapprove of the candidates via a sincere AV strategy,
  where the above-mentioned conventions apply: For elections with one
  candidate, each voter must approve of this candidate; and for
  elections $(C,V)$ with at least two candidates, each voter $v \in V$
  is required to have an AV strategy $S_v$ with $\emptyset \neq S_v
  \neq C$, which implies $||C|| \geq k+1$ for $k > 1$.
\end{definition}

} %

\subsection{Control Problems for  Preference-Based Approval Voting}
\label{sec:prelims:control-problems}

The control problems considered here were introduced by Bartholdi,
Tovey, and Trick~\cite{bar-tov-tri:j:control} for constructive control
and by Hemaspaandra, Hemaspaandra, and
Rothe~\cite{hem-hem-rot:j:destructive-control} for destructive
control.  In constructive control scenarios the chair's goal is to
make a favorite candidate win, and in destructive control scenarios
the chair's goal is to ensure that a despised candidate does not win.
As is common, the chair is assumed to have complete knowledge of the
voters' preference rankings and approval strategies,\footnote{A detailed
discussion of this assumption can be found 
in~\cite{hem-hem-rot:j:destructive-control}.  In a nutshell, one
justification of this assumption is that it is realistic in
many (though certainly not in all)
situations, particularly
in those involving small-scale private elections and in those involving
large-scale elections among software agents
that cooperate in a multiagent
environment and have an incentive to
reveal their preferences over the given alternatives.
Another justification is that this paper focuses on proving control
resistances of (i.e., $\np$-hardness results for) SP-AV, and an
$\np$-hardness result in the more restrictive setting of complete
knowledge clearly implies the corresponding $\np$-hardness result
in the more flexible setting of partial knowledge (see
\cite{hem-hem-rot:j:destructive-control} for more discussion of this point).}
and as in most papers on electoral
control we
define the control problems in the unique-winner model.\footnote{Exceptions
are, e.g., 
\cite{mei-pro-ros-zoh:j:multiwinner,fal-hem-hem-rot:j-To-Appear-With-TR-Ptr:llull-copeland-full-techreport,fal-hem-hem-rot:c:copeland-fully-resists-constructive-control,fal-hem-hem-rot:c:llull}, where \cite{fal-hem-hem-rot:j-To-Appear-With-TR-Ptr:llull-copeland-full-techreport,fal-hem-hem-rot:c:copeland-fully-resists-constructive-control,fal-hem-hem-rot:c:llull}
consider both the unique-winner model and the nonunique-winner model.}
In this model, the chair seeks to, via the control action, either make a
designated candidate the unique winner (in the constructive case) or
to prevent a designated candidate from being a unique winner (in the
destructive case).

To achieve his or her goal, the chair modifies the structure of a
given election via adding/deleting/partitioning either candidates or
voters.  Such control actions---specifically those with respect to
control via deleting or partitioning candidates or via partitioning
voters---may have an undesirable impact on the resulting
election in that they might
turn admissible AV strategies into inadmissible ones.
That is why we define the following rule that
coerces
admissibility (even under such control actions):
\begin{rules}[AV Strategy Rewrite Rule]
\label{rul:preference-rewrite-rule}
If in an election $(C,V)$ with $\|C\| \geq 2$ we have $S_v =
\emptyset$ or $S_v = C$ for some voter $v \in V$, then each such
voter's AV strategy is
adjusted to approve of his or her top candidate
and to disapprove of his or her bottom candidate.
\end{rules}

This rule
coerces $\emptyset \neq S_v \neq C$ for each $v \in V$ whenever there
are at least two candidates.  That is,
though it is legal for a voter to cast an inadmissible vote, the SP-AV
system will rewrite this vote to make it admissible.  In
Section~\ref{sec:discussion-spav} below, we will briefly discuss the
SP-AV system and, in particular, some subtle points regarding
Rule~\ref{rul:preference-rewrite-rule}.

We now formally define our control problems, where each problem is
defined by stating the problem instance together with two questions,
one for the constructive and one for the destructive case.  These
control problems are tailored to sincere-strategy preference-based
approval voting by requiring every election occurring in these control
problems (be it before, during, or after a control action---so, in
particular, this also applies to the subelections in the partitioning
cases) to have a sincere AV strategy profile.
Note that when the number
of candidates is reduced (due to deleting candidates or partitioning
candidates or voters), approval lines may have to be moved in accordance with
Rule~\ref{rul:preference-rewrite-rule}.

To avoid unnecessary repetition, when defining the $22$ control
scenarios and problems considered in this paper, we will omit (or only
very briefly sketch) the motivation of these control scenarios.  Note,
however, that each scenario considered has a natural real-world
interpretation---ranging from ``get-out-the-vote'' drives (control by
adding voters) over vote suppression or disenfranchisement (control by
deleting voters) to gerrymandering (control by partitioning voters)
for voter control, and similarly natural real-world interpretations
have been discussed in detail for the single cases of candidate
control.  These real-world interpretations and motivating examples
have been described at length in a number of previous papers on
control, such
as~\cite{bar-tov-tri:j:control,hem-hem-rot:j:destructive-control,fal-hem-hem-rot:j-To-Appear-With-TR-Ptr:llull-copeland-full-techreport,hem-hem-rot:c:hybrid}.
(Note that the journal version of \cite{hem-hem-rot:c:hybrid} appears
in the same special issue as the present paper.)

\subsubsection{Control by Adding Candidates}
\label{sec:control-by-adding-candidates}

In this control scenario, the chair seeks to reach his or her goal by
adding to the election, which originally involves only ``qualified''
candidates, some new candidates who are chosen from a given pool of
spoiler candidates.  In their study of control for plurality, Condorcet,
and approval voting,
Hemaspaandra, Hemaspaandra, and
Rothe~\cite{hem-hem-rot:j:destructive-control} considered only the
case of adding an \emph{unlimited} number of spoiler candidates (which
is the original variant of this problem as defined by Bartholdi,
Tovey, and Trick~\cite{bar-tov-tri:j:control}).  We
consider the same variant of this problem here to make our results
comparable with those established
in~\cite{hem-hem-rot:j:destructive-control}, but for completeness we
in addition consider the case of adding a \emph{limited} number of
spoiler candidates, where the prespecified limit is part of the
problem instance.  This variant of this problem was introduced by
Faliszewski et
al.~\cite{fal-hem-hem-rot:j-To-Appear-With-TR-Ptr:llull-copeland-full-techreport,fal-hem-hem-rot:c:llull,fal-hem-hem-rot:c:copeland-fully-resists-constructive-control}
in analogy with the definitions of control by deleting candidates and
of control by adding or deleting voters.  They showed that, for the
election system Copeland$^{\alpha}$ they investigate, the complexity
of these two problems can
drastically change depending on the parameter~$\alpha$,
see~\cite{fal-hem-hem-rot:j-To-Appear-With-TR-Ptr:llull-copeland-full-techreport,fal-hem-hem-rot:c:copeland-fully-resists-constructive-control}.

We first define the unlimited variant of control by adding candidates.

\begin{desctight}

\item[Name] Control by Adding an Unlimited Number of Candidates.

\item[Instance] An election $(C \cup D, V)$ and a designated
candidate $c \in C$, where the set $C$ of qualified candidates and the
set $D$ of spoiler candidates are disjoint.

\item[Question (constructive)] Is it possible to choose a subset $D'
\seq D$ such that $c$ is the unique winner of election $(C \cup D', V)$?

\item[Question (destructive)] Is it possible to choose a subset $D'
\seq D$ such that $c$ is not a unique winner of election $(C \cup D', V)$?
\end{desctight}

The problem Control by Adding a Limited Number of Candidates is
defined analogously, with the only difference being that the chair
seeks to reach his or her goal by adding at most $\ell$ spoiler
candidates, where $\ell$ is part of the problem instance.

\OMIT{
In the problem instance above, we additionally require $\|C\| \geq k$
for $k$-AV, $\|C\| \geq 1$ for SP-$1$-AV, and $\|C\| \geq k+1$ for
SP-$k$-AV with $k > 1$.
} %

\subsubsection{Control by Deleting Candidates}

In this control scenario, the chair seeks to reach his or her goal by
deleting (up to a given number of) candidates.  Here it may happen
that
inadmissible AV strategies are created by the control action, but
Rule~\ref{rul:preference-rewrite-rule} will coerce admissibility again
(by moving the line between
some voter's acceptable and inacceptable candidates to behind the top
candidate or to before the bottom candidate whenever necessary).

\begin{desctight}

\item[Name] Control by Deleting Candidates.

\item[Instance] An election $(C,V)$, a designated candidate $c \in
C$, and a nonnegative integer~$\ell$.

\item[Question (constructive)] Is it possible to delete up to $\ell$
candidates from $C$ such that $c$ is the unique winner of the
resulting election?

\item[Question (destructive)] Is it possible to delete up to $\ell$
candidates (other than~$c$) from $C$ such that $c$ is not a unique
winner of the resulting election?
\end{desctight}

\subsubsection{Control by Partition and Run-Off Partition of Candidates}

There are two partition-of-candidates control scenarios.  In both
scenarios, the chair seeks to reach his or her goal by partitioning
the candidate set $C$ into two subsets, $C_1$ and $C_2$, after which
the election is conducted in two stages.  In control by partition of
candidates, the election's first stage is held within only one group,
say~$C_1$, and this group's winners that survive the tie-handling rule
used (see the next paragraph) run against all members of~$C_2$ in the
second and final stage.  In control by run-off partition of
candidates, the election's first stage is held separately within both
groups, $C_1$ and $C_2$, and the winners of both subelections that
survive the tie-handling rule used run against each other in the
second and final stage.

We use the two tie-handling rules proposed by Hemaspaandra,
Hemaspaandra, and Rothe \cite{hem-hem-rot:j:destructive-control}:
ties-promote (TP) and ties-eliminate (TE).  In the TP model, all the
first-stage winners of a subelection, $(C_1,V)$ or $(C_2,V)$, are
promoted to the final round.  In the TE model, a first-stage winner of
a subelection, $(C_1,V)$ or $(C_2,V)$, is promoted to the final round
exactly if he or she is that subelection's unique winner.

Note that partitioning the candidate set $C$ into $C_1$ and $C_2$ is,
in some sense, similar to deleting $C_2$ from $C$ to obtain subelection
$(C_1,V)$ and to deleting $C_1$ from~$C$ to obtain subelection
$(C_2,V)$.  Also, the
final stage of the election may have a reduced number of candidates
(which depends on the tie-handling rule used).
So, in the partitioning cases, it may again happen that
inadmissible AV strategies are created by the control action, but
Rule~\ref{rul:preference-rewrite-rule} will coerce admissibility again.

\begin{desctight}

\item[Name] Control by Partition of Candidates.

\item[Instance] An election $(C,V)$ and a designated candidate $c \in
C$.

\item[Question (constructive)] Is it possible to partition $C$ into
$C_1$ and $C_2$ such that $c$ is the unique winner of the final stage
of the two-stage election in which the winners of subelection
$(C_1,V)$ that survive the tie-handling rule run against all
candidates in $C_2$ (with respect to the votes in~$V$)?

\item[Question (destructive)] Is it possible to partition $C$ into
$C_1$ and $C_2$ such that $c$ is not a unique winner of the final
stage of the two-stage election in which the winners of subelection
$(C_1,V)$ that survive the tie-handling rule run against all
candidates in $C_2$ (with respect to the votes in~$V$)?
\end{desctight}

\begin{desctight}

\item[Name] Control by Run-Off Partition of Candidates.

\item[Instance] An election $(C,V)$ and a designated candidate $c
\in C$.

\item[Question (constructive)] Is it possible to partition $C$ into
$C_1$ and $C_2$ such that $c$ is the unique winner of the final stage
of the two-stage election in which the winners of subelection
$(C_1,V)$ that survive the tie-handling rule run (with respect to the
votes in~$V$) against the winners of subelection $(C_2,V)$ that
survive the tie-handling rule?

\item[Question (destructive)] Is it possible to partition $C$ into
$C_1$ and $C_2$ such that $c$ is not a unique winner of the final
stage of the two-stage election in which the winners of subelection
$(C_1,V)$ that survive the tie-handling rule run (with respect to the
votes in~$V$) against the winners of subelection $(C_2,V)$ that
survive the tie-handling rule?
\end{desctight}

\subsubsection{Control by Adding Voters}

In this control scenario, the chair seeks to reach his or her goal by
introducing new voters into a given election.  These additional voters
are chosen from a given pool of voters whose preferences and approval
strategies over the candidates from the original election are known.
Again, the number of voters that can be added is prespecified.

\begin{desctight}

\item[Name] Control by Adding Voters.

\item[Instance] An election $(C,V)$, a collection $W$ of additional
voters with known preferences and approval strategies over~$C$, a
designated candidate $c \in C$, and a nonnegative integer~$\ell$.

\item[Question (constructive)] Is it possible to choose a subset $W'
\seq W$ with $\|W'\| \leq \ell$ such that $c$ is the unique winner of
election $(C,V \cup W')$?

\item[Question (destructive)] Is it possible to choose a subset $W'
\seq W$ with $\|W'\| \leq \ell$ such that $c$ is not a unique winner
of election $(C,V \cup W')$?
\end{desctight}

\subsubsection{Control by Deleting Voters}

The chair here seeks to reach his or her goal by suppressing (up to a
prespecified number of) voters.

\begin{desctight}

\item[Name] Control by Deleting Voters.

\item[Instance] An election $(C,V)$, a designated candidate $c \in
C$, and a nonnegative integer~$\ell$.

\item[Question (constructive)] Is it possible to delete up to $\ell$
voters from $V$ such that $c$ is the unique winner of the resulting
election?

\item[Question (destructive)] Is it possible to delete up to $\ell$
voters from $V$ such that $c$ is not a unique winner of the resulting
election?
\end{desctight}

\subsubsection{Control by Partition of Voters}

In this scenario, the election again is conducted in two stages, and
the chair now seeks to reach his or her goal by partitioning the
voters $V$ into two subcommittees, $V_1$ and~$V_2$.  In the first
stage, the subelections $(C,V_1)$ and $(C,V_2)$ are held separately in
parallel, and the winners of each subelection who survive the
tie-handling rule move forward to the second and final stage in which
they compete against each other.

As in the candidate-deletion and the candidate-partition cases, also
in control by partition of voters it may happen that inadmissible AV
strategies are created by the control action, since the final stage of
the election may have a reduced number of candidates.  However, if
that happens then Rule~\ref{rul:preference-rewrite-rule} will again
coerce admissibility.

\begin{desctight}

\item[Name] Control by Partition of Voters.

\item[Instance] An election $(C,V)$ and a designated candidate $c \in
C$.

\item[Question (constructive)] Is it possible to partition $V$ into
$V_1$ and $V_2$ such that $c$ is the unique winner of the final stage
of the two-stage election in which the winners of subelection
$(C,V_1)$ that survive the tie-handling rule run (with respect to the
votes in~$V$) against the winners of subelection $(C,V_2)$ that
survive the tie-handling rule?

\item[Question (destructive)] Is it possible to partition $V$ into
$V_1$ and $V_2$ such that $c$ is not a unique winner of the final
stage of the two-stage election in which the winners of subelection
$(C,V_1)$ that survive the tie-handling rule run (with respect to the
votes in~$V$) against the winners of subelection $(C,V_2)$ that
survive the tie-handling rule?
\end{desctight}

\subsection{A Brief Discussion of SP-AV}
\label{sec:discussion-spav}

The notion of SP-AV, as defined here, slightly differs from the
definition proposed in this paper's
precursors~\cite{erd-now-rot:c:sp-av,erd-now-rot:t:sp-av}.  For
example, \cite{erd-now-rot:c:sp-av} specifically required for
single-candidate elections that each voter must approve of this
candidate.  In the present paper, we drop this requirement,
as it in fact is not needed (because
the one candidate in a single-candidate election will always
win---even with zero approvals, i.e., SP-AV is a ``voiced'' voting
system).

The other definitional change is more subtle.
In~\cite{erd-now-rot:c:sp-av,erd-now-rot:t:sp-av}, we adopted Brams
and Sanver's
assumption that voters $v$ are required to have admissible AV
strategies (i.e., only AV strategies $S_v$ with $\emptyset \neq S_v
\neq C$ were allowed).\footnote{Except that
\cite{bra-san:j:critical-strategies-under-approval} excludes only the
case $S_v \neq C$, see Footnote~\ref{foo:admissible}.}
By this assumption, any vote with an inadmissible AV strategy was
considered
void, and we applied our rule of rewriting inadmissible AV strategies
to coerce admissibility only when a control action had turned an
originally admissible vote into an inadmissible one.  One problem with
this approach was that this rule depended on (and could be viewed as
redefining)
control rather than being an integral part of the voting system
itself.  In contrast, we
now allow voters to cast inadmissible votes, and
Rule~\ref{rul:preference-rewrite-rule} will turn them into admissible
votes the same way it will coerce admissibility for votes that became
inadmissible in the course of a control action.  The in-preparation
bookchapter~\cite{bau-erd-hem-hem-rot:btoappear-m:computational-apects-of-approval-voting}
elaborates on this point and on other points regarding the
definitional changes SP-AV has undergone in the course of its
development up to its final form in the present paper.  We stress that
none of the two changes mentioned above has a severe impact on our
findings or their proofs.

Another issue to be addressed is that the choice of
Rule~\ref{rul:preference-rewrite-rule} might seem to be purely a
matter of taste, at first glance.  For example, given an inadmissible
AV strategy of the form $\begin{array}{@{}c@{\ \ }c@{\ \ }c@{\ \ }c@{\
\ }c@{}} | & a & b & c & d \end{array}$ (respectively,
$\begin{array}{@{}c@{\ \ }c@{\ \ }c@{\ \ }c@{\ \ }c@{}} a & b & c & d
& | \end{array}$), why don't we change it into an admissible vote of
the form, say, $\begin{array}{@{}c@{\ \ }c@{\ \ }c@{\ \ }c@{\ \ }c@{}}
a & b & | & c & d \end{array}$ rather than, according to
Rule~\ref{rul:preference-rewrite-rule}, into $\begin{array}{@{}c@{\ \
}c@{\ \ }c@{\ \ }c@{\ \ }c@{}} a & | & b & c & d \end{array}$
(respectively, $\begin{array}{@{}c@{\ \ }c@{\ \ }c@{\ \ }c@{\ \ }c@{}}
a & b & c & | & d \end{array}$)?  The reason is that, once we have
agreed that it is desirable to coerce admissibility, our choice of
Rule~\ref{rul:preference-rewrite-rule} is the most sensible way, as
this is the minimally invasive rule to coerce admissibility among all
possible such rules: We do change the voters' approval strategies, but
we wish to do this in the least harmful way.

\subsection{Immunity, Susceptibility, Vulnerability, and Resistance}
\label{sec:prelims:immunity-susceptibility-vulnerability-resistance}

The following notions---which are due to Bartholdi, Tovey, and
Trick~\cite{bar-tov-tri:j:control} (see also
\cite{hem-hem-rot:j:destructive-control,hem-hem-rot:c:hybrid,fal-hem-hem-rot:j-To-Appear-With-TR-Ptr:llull-copeland-full-techreport,fal-hem-hem-rot:c:copeland-fully-resists-constructive-control,fal-hem-hem-rot:c:llull})---will 
be central to our complexity analysis of the control problems for
{SP-AV}.

\begin{definition}
Let $\mathcal{E}$ be an election system and let $\Phi$ be some given
type of control.
\begin{enumerate}
\item $\mathcal{E}$ is said to be \emph{immune to $\Phi$-control} if
\begin{enumerate}
\item $\Phi$ is a constructive control type and it is never possible
for the chair to turn a designated candidate from being not a unique
winner into being the unique winner via exerting $\Phi$-control, or
\item $\Phi$ is a destructive control type and it is never possible for
the chair to turn a designated candidate from being the unique winner
into being not a unique winner via exerting $\Phi$-control.
\end{enumerate}

\item $\mathcal{E}$ is said to be \emph{susceptible to $\Phi$-control}
if it is not immune to $\Phi$-control.

\item $\mathcal{E}$ is said to be \emph{vulnerable to $\Phi$-control}
if $\mathcal{E}$ is susceptible to $\Phi$-control and the control
problem associated with $\Phi$ is solvable in polynomial time.

\item $\mathcal{E}$ is said to be \emph{resistant to $\Phi$-control}
if $\mathcal{E}$ is susceptible to $\Phi$-control and the control
problem associated with $\Phi$ is $\np$-hard.
\end{enumerate}
\end{definition}

For example, approval voting is known to be immune to eight of the
twelve types of candidate
control considered
in~\cite{hem-hem-rot:j:destructive-control}.  The proofs of these
results crucially employ the
equivalences and implications between immunity/susceptibility for
various control types shown
in~\cite{hem-hem-rot:j:destructive-control} and the fact that approval
voting satisfies the unique version of the Weak Axiom of Revealed
Preference (denoted by Unique-WARP,
see~\cite{hem-hem-rot:j:destructive-control,bar-tov-tri:j:control}):
If a candidate $c$ is the unique winner in a set $C$ of candidates,
then $c$ is the unique winner in every subset of $C$ that
includes~$c$.  In contrast with approval voting, sincere-strategy
preference-based approval voting does not satisfy Unique-WARP, and we
will see later in Section~\ref{sec:results:susceptibility} that it
indeed is susceptible to each type of control considered here.

\begin{proposition}
\label{prop:unique-warp}
Sincere-strategy preference-based approval voting does not satisfy
Unique-WARP.
\end{proposition}

\begin{proofs}
  Consider the election $(C,V)$ with candidate set $C = \{a, b, c,
  d\}$ and voter collection $V = \{v_1, v_2, v_3, v_4\}$.  Removing
  candidate $d$ changes the profile as follows according to
  Rule~\ref{rul:preference-rewrite-rule}:
\[
\begin{array}{lc@{\ \ }c@{\ \ }c@{\ \ }c@{\ \ }c@{\quad\quad}c@{\quad\quad}c@{\ \ }c@{\ \ }c@{\ \ }c}
v_1: & b & c & a & | & d &                           & b & c & | & a \\
v_2: & c & | & a & d & b & \mbox{is changed to}      & c & | & a & b \\
v_3: & a & b & c & | & d & \mbox{(by removing $d$):} & a & b & | & c \\
v_4: & b & a & c & | & d &                           & b & a & | & c \\
\end{array}
\]

Note that the approval/disapproval line has been moved in voters
$v_1$, $v_3$, and~$v_4$
according to Rule~\ref{rul:preference-rewrite-rule}.
Although $c$ was the unique winner of
$(C,V)$, $c$ is not a winner of $(\{a,b,c\},V)$ (in fact, $b$ is the
unique winner of $(\{a,b,c\},V)$).  Thus, SP-AV does not satisfy
Unique-WARP.~\end{proofs}

\section{Results for Sincere-Strategy Preference-Based Approval Voting}
\label{sec:results}

Theorem~\ref{thm:summary-of-results} below (see also
Table~\ref{tab:summary-of-results}) shows the complexity results
regarding control of elections for
{SP-AV}.  As mentioned in the introduction, with $19$
resistances and only three vulnerabilities, this system has more
resistances and fewer vulnerabilities to control (for our $22$ control
types) than is currently known for any other natural voting system
with a polynomial-time winner problem.

\begin{theorem}
\label{thm:summary-of-results}
Sincere-strategy preference-based approval voting has the resistances and
vulnerabilities to the $22$ types of control defined in
Section~\ref{sec:prelims:control-problems} that are shown in
Table~\ref{tab:summary-of-results}.
\end{theorem}

\begin{table}[t+]
\centering
{\scriptsize
\begin{tabular}{|l||l|l||l|l||l|l|}
\hline
                    & \multicolumn{2}{c||}{Plurality}
                    & \multicolumn{2}{c||}{SP-AV}
                    & \multicolumn{2}{c|}{AV}
\\ \cline{2-7}
Control by          & Constr. & Destr.
                    & Constr. & Destr.
                    & Constr. & Destr.
\\ \hline\hline
Adding an Unlimited Number of Candidates
                    &      R                &      R 
                    & {\bf R}               & {\bf R} 
                    &      I                &      V 
\\ \hline
Adding a Limited Number of Candidates
                    &      R                &      R 
                    & {\bf R}               & {\bf R} 
                    &      I                &      V 
\\ \hline
Deleting Candidates &      R                &      R 
                    & {\bf R}               & {\bf R} 
                    &      V                &      I 
\\ \hline
Partition of Candidates
                    &      TE: R            &      TE: R 
                    & {\bf TE: R}           & {\bf TE: R} 
                    &      TE: V            &      TE: I 
\\
                    &      TP: R            &      TP: R 
                    & {\bf TP: R}           & {\bf TP: R}      
                    &      TP: I            &      TP: I 
\\ \hline
Run-off Partition of Candidates
                    &      TE: R            &      TE: R 
                    &      {\bf TE: R}      &      {\bf TE: R} 
                    &      TE: V            &      TE: I 
\\ 
                    &      TP: R            &      TP: R 
                    & {\bf TP: R}           & {\bf TP: R}      
                    &      TP: I            &      TP: I 
\\ \hline
Adding Voters       &      V                &      V  
                    & {\bf R}               & {\bf V} 
                    &      R                &      V 
\\ \hline
Deleting Voters     &      V                &      V 
                    & {\bf R}               & {\bf V} 
                    &      R                &      V 
\\ \hline
Partition of Voters
                    &      TE: V            &      TE: V 
                    & {\bf TE: R}           & {\bf TE: V} 
                    &      TE: R            &      TE: V 
\\
                    &      TP: R            &      TP: R 
                    & {\bf TP: R}           & {\bf TP: R}      
                    &      TP: R            &      TP: V 
\\ 
\hline
\end{tabular}
}
\caption{\label{tab:summary-of-results}
Overview of results.  Key: I means immune,
R means resistant,
V means vulnerable,
TE means ties-eliminate, and TP means ties-promote.
Results for SP-AV are new; their proofs are either new or draw on
proofs from~\cite{hem-hem-rot:j:destructive-control}.
Results for plurality and AV,
stated here to allow comparison, are due to Bartholdi, Tovey, and
Trick~\cite{bar-tov-tri:j:control} and to Hemaspaandra,
Hemaspaandra, and Rothe~\cite{hem-hem-rot:j:destructive-control}.
(The results for control by adding a limited number of candidates
for plurality and approval voting, though not stated explicitly
in~\cite{bar-tov-tri:j:control,hem-hem-rot:j:destructive-control},
follow immediately from
the proofs of the corresponding results for the ``unlimited'' variant
of the problem, see Footnote~\ref{foo:plurality}.)}
\end{table}

\subsection{Susceptibility}
\label{sec:results:susceptibility}

By definition, all resistance and vulnerability results in particular
require susceptibility.  In the following two lemmas, we prove that
sincere-strategy preference-based approval voting is susceptible to
the $22$ types of control defined in
Section~\ref{sec:prelims:control-problems}.  To this end, we will make
use of Theorems~4.1, 4.2, and 4.3 of Hemaspaandra, Hemaspaandra, and
Rothe~\cite{hem-hem-rot:j:destructive-control} that provide
susceptibility
equivalences and implications for various control types.\footnote{Although
\cite{hem-hem-rot:j:destructive-control} does not consider
the case of control by adding a limited number of candidates
explicitly, it is immediate that all proofs for the ``unlimited'' case
in \cite{hem-hem-rot:j:destructive-control} work also for this
``limited'' case.}
\jfootnote{In der Journalversion sollten wir Theorems~4.1, 4.2 und 4.3 aus
\cite{hem-hem-rot:j:destructive-control} explizit angeben.  Da die
Susceptibility-Aussagen in der Konferenzversion nur im Anhang
auftauchen, waehle ich aus Platzgruenden diese Kurzvariante.}
For the sake of self-containment, we give these results below,
stated essentially 
word-for-word as in~\cite{hem-hem-rot:j:destructive-control}.
In particular, Theorem~\ref{thm:two-way}
(which is \cite[Thm.~4.1]{hem-hem-rot:j:destructive-control}) gives four
equivalences between susceptibility to constructive/destructive
control by adding/deleting candidates/voters;
Theorem~\ref{thm:one-way} 
(which is \cite[Thm.~4.2]{hem-hem-rot:j:destructive-control}) gives four
implications that link susceptibility to control by (run-off)
partition of candidates/voters with susceptibility to control by
deleting candidates/voters; and
Theorem~\ref{thm:hinged}
(which is \cite[Thm.~4.3]{hem-hem-rot:j:destructive-control})
states that every
``voiced'' voting system is susceptible to constructive control by
deleting candidates and to destructive control by adding candidates,
and that for each voiced voting system susceptibility to destructive
control by partition of voters (in model TE or TP) implies
susceptibility to destructive control by deleting voters.  A voting
system is said to be \emph{voiced} if in every one-candidate election,
this candidate wins.

\begin{theorem}[Thm.~4.1 of \cite{hem-hem-rot:j:destructive-control}]
\label{thm:two-way}
\begin{enumerate}
\item A voting system is 
susceptible to constructive control
by adding (either a limited or an unlimited number of)
candidates if and only if it is 
susceptible to destructive control
by deleting candidates.
\item \label{part:two-way:two} A voting system is 
susceptible to constructive control
by deleting candidates if and only if it is 
susceptible to destructive control
by adding (either a limited or an unlimited number of) candidates.

\item A voting system is 
susceptible to constructive control
by adding voters if and only if it is 
susceptible to destructive control
by deleting voters.
\item A voting system is 
susceptible to constructive control
by deleting voters if and only if it is 
susceptible to destructive control
by adding voters.
\end{enumerate}
\end{theorem}

\begin{theorem}[Thm.~4.2 of \cite{hem-hem-rot:j:destructive-control}]
\label{thm:one-way}
\begin{enumerate}
\item \label{part:one-way:surprise} If a voting system is 
susceptible to constructive control
by partition of voters 
(in model TE or TP), then it is 
susceptible to constructive control
by deleting 
candidates.

\item \label{part:one-way:boring} If a voting system is 
susceptible to constructive control
by partition or run-off partition of candidates
(in model TE or TP), then it is 
susceptible to constructive control
by deleting candidates.

\item \label{part:one-way:three} If a voting system is 
susceptible to constructive control
by partition of voters 
in model TE, then it is 
susceptible to constructive control
by deleting 
voters.

\item \label{part:one-way:four} If a voting system is 
susceptible to destructive control
by partition or run-off partition of candidates
(in model TE or TP), then it is 
susceptible to destructive control
by deleting candidates.
\end{enumerate}
\end{theorem}

\begin{theorem}[Thm.~4.3 of \cite{hem-hem-rot:j:destructive-control}]
\label{thm:hinged}
\begin{enumerate}
\item \label{part:hinged:one} If a voiced voting system is 
susceptible to destructive control
by partition of voters 
(in model TE or TP), then it is 
susceptible to destructive control
by deleting 
voters.

\item \label{part:hinged:two} Each voiced voting system is 
susceptible to constructive control
by deleting 
candidates.

\item \label{part:hinged:three} Each voiced voting system is 
susceptible to destructive control
by adding (either a limited or an unlimited number of) candidates.
\end{enumerate}
\end{theorem}

We start with susceptibility to candidate control for SP-AV.

\begin{lemma}
\label{lem:susceptible-candidate-control}
SP-AV is susceptible to constructive and destructive control by adding
candidates (in both the ``limited'' and the ``unlimited'' variant of the
problem), by deleting candidates, and by partition of candidates
(with or without run-off and for each in both tie-handling models, TE and TP).
\end{lemma}

\begin{proofs}
From
Theorem~\ref{thm:hinged} and the
obvious fact that SP-AV is a voiced voting system, it immediately
follows that SP-AV is susceptible to constructive control by deleting
candidates and to destructive control by adding candidates (in both
the ``limited'' and the ``unlimited'' variant of the problem).

Consider the election $(C,V)$ with candidate set $C = \{a, b, c,
d, e, f\}$ and voter collection $V = \{v_1, v_2, \ldots , v_6\}$ and
the following partition of $C$ into $C_1 = \{a,c,d\}$ and $C_2 =
\{b,e,f\}$:
\[
\begin{array}{lc@{\ \ }c@{\ \ }c@{\ \ }c@{\ \ }c@{\ \ }c@{\ \ }c@{\ \ }c@{\ \ }c@{\ \ }c@{\ \ }c@{\ \ }c@{\ \ }c@{\ \ }c@{\ \ }c@{\ \ }c@{\ \ }c}
 & \multicolumn{7}{c}{(C,V)} & \mbox{ is partitioned into }
 & \multicolumn{4}{c}{(C_1,V)} & \mbox{ and }
 & \multicolumn{4}{c}{(C_2,V)}
\\ \cline{2-8}\cline{10-13}\cline{15-18}
v_1: & a & b & c & | & d & e & f &   & a & c & | & d &   & b & | & e & f \\
v_2: & b & c & | & a & d & e & f &   & c & | & a & d &   & b & | & e & f \\
v_3: & a & c & | & b & d & e & f &   & a & c & | & d &   & b & | & e & f \\
v_4: & b & a & c & | & d & e & f &   & a & c & | & d &   & b & | & e & f \\
v_5: & a & b & d & e & c & | & f &   & a & d & | & c &   & b & e & | & f \\
v_6: & a & e & d & f & c & | & b &   & a & d & | & c &   & e & f & | & b \\
\end{array}
\]
With six approvals, $c$ is the unique winner of $(C,V)$.  However, $a$
is the unique winner of $(C_1,V)$, which implies that $c$ is not
promoted to the final stage, regardless of whether we use the TE or TP
tie-handling rule and regardless of whether we employ a partition of
candidates with or without run-off.  Thus, SP-AV is susceptible to
destructive control by partition of candidates (with or without
run-off and for each in both tie-handling models, TE and TP).  By
Theorem~\ref{thm:one-way}, SP-AV is also
susceptible to destructive control by deleting candidates.  By
Theorem~\ref{thm:two-way}
in turn, SP-AV is
also susceptible to constructive control by adding
candidates (in both the ``limited'' and the ``unlimited'' 
variant of the problem).

Note that $a$ is not the unique winner of $(C,V)$, as $a$ loses to
$c$ by $5$ to~$6$.  However, if we partition $C$ into $C_1 =
\{a,c,d\}$ and $C_2 = \{b,e,f\}$, then $a$ is the unique winner of
$(C_1,V)$ and $b$ is the unique winner of $(C_2,V)$.  Since both
subelections have a unique winner, it does not matter whether the TE
rule or the TP rule is applied.  The final-stage election is
$(\{a,b\},V)$ in the case of run-off partition of candidates, and it
is $(\{a,b,e,f\},V)$ in the case of partition of candidates.  Since
$a$ wins against $b$ in the former case by $4$ to $2$ and in the
latter case by $5$ to $4$ (and $e$ and $f$ do even worse than $b$ in
this case), $a$ is the unique winner in both cases.  Thus, SP-AV is
susceptible to constructive control by partition of candidates (with
or without run-off and for each in both models, TE and TP).~\end{proofs}

We now turn to susceptibility to voter control.

\begin{lemma}
\label{lem:susceptible-voter-control}
SP-AV is susceptible to constructive and destructive control by adding
voters, by deleting voters, and by partition of voters in both
tie-handling models, TE and~{TP}.
\end{lemma}

\begin{proofs}
Consider the election $(C,V)$ with candidate set $C = \{a, b, c, d, e,
f\}$ and voter collection $V = \{v_1, v_2, \ldots , v_8\}$ and
partition $V$ into $V_1 = \{v_1,v_2,v_3,v_4\}$ and $V_2 =
\{v_5,v_6,v_7,v_8\}$.  Thus, we change:
\[
\begin{array}{lc@{\ \ }c@{\ \ }c@{\ \ }c@{\ \ }c@{\ \ }c@{\ \ }c@{\ \ }c@{\ \ }c@{\ \ }c@{\ \ }c@{\ \ }c@{\ \ }c@{\ \ }c@{\ \ }c@{\ \ }c@{\ \ }c@{\ \ }c@{\ \ }c@{\ \ }c@{\ \ }c@{\ \ }c@{\ \ }c}
 & \multicolumn{7}{c}{(C,V)} & \mbox{ into }
 & \multicolumn{7}{c}{(C,V_1)} & \mbox{ and }
 & \multicolumn{7}{c}{(C,V_2)}
\\ \cline{2-8}\cline{10-16}\cline{18-24}
v_1: & a & b & c & | & d & e & f &
     & a & b & c & | & d & e & f &
     &   &   &   &   &   &   &   \\
v_2: & a & c & | & b & d & e & f &
     & a & c & | & b & d & e & f &
     &   &   &   &   &   &   &   \\
v_3: & c & b & a & d & | & e & f &
     & c & b & a & d & | & e & f &
     &   &   &   &   &   &   &   \\
v_4: & a & b & | & d & e & c & f &
     & a & b & | & d & e & c & f &
     &   &   &   &   &   &   &   \\
v_5: & a & d & c & | & b & e & f &
     &   &   &   &   &   &   &   &
     & a & d & c & | & b & e & f \\
v_6: & e & b & c & d & | & a & f &
     &   &   &   &   &   &   &   &
     & e & b & c & d & | & a & f \\
v_7: & d & e & c & f & | & b & a &
     &   &   &   &   &   &   &   &
     & d & e & c & f & | & b & a \\
v_8: & d & f & | & b & a & c & e &
     &   &   &   &   &   &   &   &
     & d & f & | & b & a & c & e \\
\end{array}
\]
With six approvals, $c$ is the unique winner of $(C,V)$.  However, $a$
is the unique winner of $(C,V_1)$ and $d$ is the unique winner of
$(C,V_2)$, which implies that $c$ is not promoted to the final stage,
regardless of whether we use the TE or TP tie-handling rule.  (In the
final-stage election $(\{a,d\},V)$, $a$ wins by $5$ to~$3$.)  Thus,
SP-AV is susceptible to destructive control by partition of voters in
models TE and {TP}.  By
Theorem~\ref{thm:hinged} and since SP-AV is
a voiced system, SP-AV is also susceptible to destructive control by
deleting voters.  Finally, by
Theorem~\ref{thm:two-way}, SP-AV is also
susceptible to constructive control by adding voters.

Now, if we let $a$ and $c$ change their roles in the above election
and argument, we see that SP-AV is also susceptible to constructive
control by partition of voters in models TE and {TP}.  By
Theorem~\ref{thm:one-way}, susceptibility to
constructive control by partition of voters in model TE implies
susceptibility to constructive control by deleting voters.  Again, by
Theorem~\ref{thm:two-way}, SP-AV is also
susceptible to destructive control by adding 
voters.~\end{proofs}

\subsection{Candidate Control}
\label{sec:results:candidate-control}

Theorems~\ref{thm:hem-hem-rot-candidate-control} and
\ref{thm:resistance-constructive-candidate-deleting} 
below show
that sincere-strategy preference-based approval voting is fully
resistant to candidate control.  This result should be contrasted with
that of Hemaspaandra, Hemaspaandra, and
Rothe~\cite{hem-hem-rot:j:destructive-control}, who proved immunity
and vulnerability for all cases of candidate control within approval
voting (see Table~\ref{tab:summary-of-results}).  In fact, SP-AV has
the same resistances to candidate control as plurality, and we will
show that the construction presented
in~\cite{hem-hem-rot:j:destructive-control} to prove plurality
resistant also works for sincere-strategy preference-based approval
voting in all cases of candidate control except one---namely, except
for constructive control by deleting candidates.
Theorem~\ref{thm:resistance-constructive-candidate-deleting} establishes
resistance for this one missing case.

All resistance results in this section follow via a reduction from the
$\np$-complete problem Hitting Set (see, e.g., Garey and
Johnson~\cite{gar-joh:b:int}): 
\begin{desctight}

\item[Name] Hitting Set.

\item[Instance] A set $B = \{b_1, b_2, \ldots , b_m\}$, a nonempty
collection $\mathcal{S} = \{S_1, S_2, \ldots , S_n\}$ of subsets $S_i
\seq B$,\footnote{Our assumption
that $\mathcal{S}$ be nonempty (i.e., that $n \geq 1$) is not
explicitly specified in Garey and Johnson~\cite{gar-joh:b:int}.
However, it is clear that requiring $n \geq 1$ does not change the
complexity of the problem.}
and a positive integer~$k \leq m$.

\item[Question] Does $\mathcal{S}$ have a hitting set of size at
most~$k$, i.e., is there a set $B' \seq B$ with $\|B'\| \leq k$ such
that for each~$i$, $S_i \cap B' \neq \emptyset$?
\end{desctight}

Note that some of our proofs for SP-AV are based on constructions
presented in~\cite{hem-hem-rot:j:destructive-control} to
prove the corresponding results for approval voting or plurality,
whereas some other of our results require new insights to make the
proof work for {SP-AV}.  For completeness, we will present each
construction here (even if the modification of a previous construction
is rather straightforward), explicitly stating whether it is based on
a previous construction from~\cite{hem-hem-rot:j:destructive-control},
and if so, we will state in each case on which construction it is
based and what the differences to the related previous construction
are.

\begin{theorem}
\label{thm:hem-hem-rot-candidate-control}
SP-AV is resistant to
all types of constructive and destructive candidate control defined in
Section~\ref{sec:prelims:control-problems} except for 
constructive control by deleting candidates (which will be handled
separately in
Theorem~\ref{thm:resistance-constructive-candidate-deleting}).
\end{theorem}

Resistance of SP-AV to constructive control by deleting candidates,
which is the missing case in
Theorem~\ref{thm:hem-hem-rot-candidate-control}, will be shown as
Theorem~\ref{thm:resistance-constructive-candidate-deleting} below.

The proof of Theorem~\ref{thm:hem-hem-rot-candidate-control}
is based on a construction for plurality
in~\cite{hem-hem-rot:j:destructive-control}, except that only the
arguments for \emph{destructive} candidate control are given there
(simply because plurality was shown resistant to all cases of
constructive candidate control already by Bartholdi, Tovey, and
Trick~\cite{bar-tov-tri:j:control} via different constructions).  We
now provide a short proof sketch of
Theorem~\ref{thm:hem-hem-rot-candidate-control} and the construction
from \cite{hem-hem-rot:j:destructive-control} (slightly modified so as
to formally conform with the SP-AV voter representation) in order
to (i)~show that the same construction can be used to establish all
but one resistances of SP-AV to \emph{constructive} candidate control,
and (ii)~explain why constructive control by deleting candidates
(which is missing in Theorem~\ref{thm:hem-hem-rot-candidate-control})
does \emph{not} follow from this construction.

\medskip

\sproofsketchof{Theorem~\ref{thm:hem-hem-rot-candidate-control}}
Susceptibility holds by Lemma~\ref{lem:susceptible-candidate-control}
in each case.  The resistance proofs are based on a reduction from
Hitting Set and employ
Construction~\ref{con:resistance-general-candidate-control} below,
slightly modified so as to formally conform with the SP-AV voter
representation.

\begin{construction}[Hemaspaandra et 
al.~\cite{hem-hem-rot:j:destructive-control}]
\label{con:resistance-general-candidate-control}
Let $(B,\mathcal{S},k)$ be a given instance of Hitting Set, where $B =
\{b_1, b_2, \ldots , b_m\}$ is a set, $\mathcal{S} = \{S_1, S_2,
\ldots , S_n\}$ is a nonempty collection of subsets $S_i \seq B$, and $k \leq
m$ is a positive integer.  Define the election $(C,V)$, where $C = B
\cup \{c,w\}$ is the candidate set and where $V$ consists of the
following voters:
\begin{enumerate}
\item There are $2(m-k) + 2n(k+1) + 4$ voters of the form:
$\begin{array}{c@{\ \ }c@{\ \ }c@{\ \ }c}
c & | & w & B .
\end{array}$
\item There are $2n(k+1) + 5$ voters of the form:
$\begin{array}{c@{\ \ }c@{\ \ }c@{\ \ }c}
w & | & c & B .
\end{array}$
\item For each~$i$, $1 \leq i \leq n$, there are $2(k+1)$ voters of
the form:
$\begin{array}{c@{\ \ }c@{\ \ }c@{\ \ }c@{\ \ }c}
S_i & | & c & w & (B-S_i) .
\end{array}$
\item For each~$j$, $1 \leq j \leq m$, there are two voters
of the form:
$\begin{array}{c@{\ \ }c@{\ \ }c@{\ \ }c@{\ \ }c}
b_j & | & w & c & (B-\{b_j\}) .
\end{array}$
\end{enumerate}
\end{construction}

Since
\begin{eqnarray*}
\lefteqn{\scoresub{(\{c,w\},V)}{c} - \scoresub{(\{c,w\},V)}{w}} \\
 & = &  (2(m-k)+2n(k+1)+4+2n(k+1))-(2n(k+1)+5+2m)\\
 & = &  2k(n-1)+2n-1
\end{eqnarray*}
is positive (because of $n \geq 1$), $c$ is the unique
winner of election $(\{c,w\},V)$.  The key observation is the
following proposition, which can be proven as
in~\cite{hem-hem-rot:j:destructive-control}.

\begin{proposition}[Hemaspaandra et 
al.~\cite{hem-hem-rot:j:destructive-control}]
\label{cla:resistance-candidate-control}
\begin{enumerate}
\item If $\mathcal{S}$ has a hitting set $B'$ of size~$k$, then $w$ is
the unique SP-AV winner of election $(B' \cup \{c,w\},V)$.
\item Let $D \subseteq B \cup \{w\}$.  If $c$ is not the unique SP-AV
winner of election $(D \cup \{c\},V)$, then there exists a set $B'
\seq B$ such that
\begin{enumerate}
\item $D = B' \cup \{w\}$, 
\item $w$ is the unique SP-AV winner of election $(B' \cup
  \{c,w\},V)$, and
\item $B'$ is a hitting set of $\mathcal{S}$ of size less than or
  equal to~$k$.
\end{enumerate}
\end{enumerate}
\end{proposition}

As an example, the resistance of SP-AV to constructive and destructive
control by adding  
candidates (both in the limited and the unlimited version of the problem)
now follows immediately from
Proposition~\ref{cla:resistance-candidate-control}, via mapping the
Hitting Set instance $(B,\mathcal{S},k)$ to the set $\{c,w\}$ of
qualified candidates and the set $B$ of spoiler candidates, to the
voter collection~$V$, and by having $c$ be the designated candidate in
the destructive case and by having $w$ be the designated candidate in
the constructive case.

The other cases of Theorem~\ref{thm:hem-hem-rot-candidate-control} can
be proven
similarly.~\eproofof{Theorem~\ref{thm:hem-hem-rot-candidate-control}}

Turning now to the missing case mentioned in 
Theorem~\ref{thm:hem-hem-rot-candidate-control} above: Why does
Construction~\ref{con:resistance-general-candidate-control} not work
for constructive control by deleting candidates?  Informally put, the
reason is that $c$ is the only serious rival of $w$ in the election
$(C,V)$ of
Construction~\ref{con:resistance-general-candidate-control}, so by
simply deleting $c$ the chair could make $w$ the unique SP-AV winner,
regardless of whether $\mathcal{S}$ has a hitting set of size~$k$.
However, via a different construction,
we can prove resistance also in this case.

\begin{theorem}
\label{thm:resistance-constructive-candidate-deleting}
SP-AV is resistant to constructive control by deleting candidates.
\end{theorem}

\begin{proofs}
\OMIT{
Susceptibility holds by Lemma~\ref{lem:susceptible-candidate-control}.
To prove resistance, we provide a reduction from the
$\np$-complete problem Exact Cover by Three-Sets (denoted by X3C; see,
e.g., Garey and Johnson~\cite{gar-joh:b:int}), which is defined as
follows: 
[X3C geloescht, kommt spaeter...]

Let $(B,\mathcal{S})$ be a given instance of Hitting Set, where $B =
\{b_1, b_2, \ldots , b_{3m}\}$ is a set, $\mathcal{S} = \{S_1, S_2,
\ldots , S_n\}$ is a collection of subsets $S_i \seq B$ with $\|S_i\|
= 3$ for each~$i$, $1 \leq i \leq n$.  Without loss of generality, we
may assume that $\bigcup_{i=1}^{n}S_i = B$, and we name the elements
of the $S_i$ as follows: $S_i = \{b_{i}^{1}, b_{i}^{2}, b_{i}^{3},
\}$.  We will use $b_j$ and $b_{i}^{k}$ interchangeably for the
elements of~$B$.

Define the election $(C,V)$ as follows.  $C = A \cup B \cup \{c,w\}
\cup S$ is the candidate set, where $A = \{a_1, a_2, \ldots, a_m\}$,
$S = \{s_1, s_2, \ldots , s_n\}$, and each $s_i$ in $S$ corresponds to
the set $S_i$ in~$\mathcal{S}$.  Voter collection $V$ consists of the
following $4n + 3k^2 - 5k - 1$ voters:
\begin{enumerate}
\item For each~$i$, $1 \leq i \leq n$, there is
\begin{enumerate}
\item one voter of the form
$\begin{array}{c@{\ \ }c@{\ \ }c@{\ \ }c@{\ \ }c@{\ \ }c@{\ \ }c}
s_i & | & c & A & B & (S-\{s_i\}) & w ;
\end{array}$
\item one voter of the form
$\begin{array}{c@{\ \ }c@{\ \ }c@{\ \ }c@{\ \ }c@{\ \ }c@{\ \ }c@{\ \ }c}
s_i & | & b_{i}^{1} & A & (B-\{b_{i}^{1}\}) & (S-\{s_i\}) & c & w ;
\end{array}$
\item one voter of the form
$\begin{array}{c@{\ \ }c@{\ \ }c@{\ \ }c@{\ \ }c@{\ \ }c@{\ \ }c@{\ \ }c}
s_i & | & b_{i}^{2} & A & (B-\{b_{2}^{1}\}) & (S-\{s_i\}) & c & w ;
\end{array}$
\item one voter of the form
$\begin{array}{c@{\ \ }c@{\ \ }c@{\ \ }c@{\ \ }c@{\ \ }c@{\ \ }c@{\ \ }c}
s_i & | & b_{i}^{3} & A & (B-\{b_{3}^{1}\}) & (S-\{s_i\}) & c & w .
\end{array}$
\end{enumerate}

\item There are $k-1$ voters of the form
$\begin{array}{c@{\ \ }c@{\ \ }c@{\ \ }c@{\ \ }c@{\ \ }c}
w & | & A & B & c & S .
\end{array}$

\item For each~$j$, $1 \leq j \leq 3m$, there are $k-2$ voters of the form
$\begin{array}{c@{\ \ }c@{\ \ }c@{\ \ }c@{\ \ }c@{\ \ }c@{\ \ }c}
b_j & | & A & (B-\{b_j\}) & c & S & w .
\end{array}$
\end{enumerate}
} %
Susceptibility holds by Lemma~\ref{lem:susceptible-candidate-control}.
To prove resistance, we provide a reduction from Hitting
Set.\footnote{In contrast, Bartholdi, Tovey, and
Trick~\cite{bar-tov-tri:j:control}
gave a reduction from Exact Cover by Three-Sets 
(which is defined in the proof of 
Theorem~\ref{thm:adding-deleting-voters}) to prove that plurality
is resistant to constructive control by deleting candidates.}
Let $(B,\mathcal{S},k)$ be a given instance of Hitting Set, where 
$B = \{b_1, b_2, \ldots , b_m\}$ is a set, $\mathcal{S} = \{S_1, S_2,
\ldots , S_n\}$ is a nonempty collection of subsets $S_i \seq B$, and $k < m$
is a positive integer.\footnote{Note that if
  $k = m$ then $B$ is always a hitting set of size at most $k$
  (provided that $\mathcal{S}$ contains only nonempty sets---a
  requirement that doesn't affect the $\np$-completeness of the
  problem), and we thus may require that $k < m$.
\label{foo:assumption-on-hitting-set-instance}}

Define the election $(C,V)$, where $C = B \cup \{w\}$ is the candidate
set and $V$ is the collection of voters.  We assume that the
candidates in $B$ are in an arbitrary but fixed order, and for each voter
below, this order is also used in each subset of~$B$.  For example, if
$B = \{b_1, b_2, b_3, b_4\}$ (where the elements of $B$ are ordered as
$b_1, b_2, b_3, b_4$) and some subset $S_i = \{b_1, b_3\}$ of
$B$ occurs in some voter then this voter prefers $b_1$ to $b_3$, and
so does any other voter whose preference list contains~$S_i$.

$V$ consists of the following $4n(k+1)+4m-2k+3$ voters:
\begin{enumerate}
\item For each~$i$, $1 \leq i \leq n$, there are $2(k+1)$ voters of
  the form:
$\begin{array}{c@{\ \ }c@{\ \ }c@{\ \ }c}
S_i & | & (B-S_i) & w .
\end{array}$
\item For each~$i$, $1 \leq i \leq n$, there are $2(k+1)$ voters of
  the form:
$\begin{array}{c@{\ \ }c@{\ \ }c@{\ \ }c}
(B-S_i) & w & | & S_i .
\end{array}$
\item For each~$j$, $1 \leq j \leq m$, there are two voters
of the form:
$\begin{array}{c@{\ \ }c@{\ \ }c@{\ \ }c}
b_j & | & w & (B-\{b_j\}) .
\end{array}$
\item There are $2(m-k)$ voters of the form:
$\begin{array}{c@{\ \ }c@{\ \ }c}
B & | & w .
\end{array}$
\item There are three voters of the form:
$\begin{array}{c@{\ \ }c@{\ \ }c}
w & | & B .
\end{array}$
\end{enumerate}

Since for each $b_j \in B$, the difference
\[
\scoresub{(C,V)}{w} - \scoresub{(C,V)}{b_j} = 
2n(k+1)+3 - (2n(k+1)+2+2(m-k)) = 1 - 2(m-k)
\]
is negative (due to $k<m$), $w$ loses to each member of $B$ and so
does not win election $(C,V)$.

We claim that $\mathcal{S}$ has a hitting set $B'$ of size $k$ if and
only if $w$ can be made the unique SP-AV winner by deleting at most
$m-k$ candidates.

From left to right: Suppose $\mathcal{S}$ has a hitting set $B'$ of
size~$k$.  Then, for each $b_j \in B'$,
\[
\scoresub{(B' \cup \{w\},V)}{w} - \scoresub{(B' \cup \{w\},V)}{b_j} = 
2n(k+1)+2(m-k)+3 - (2n(k+1)+2+2(m-k)) = 1,
\]
since the approval line is moved for $2(m-k)$ voters of the third
group according to Rule~\ref{rul:preference-rewrite-rule},
thus transferring their approvals from members of $B-B'$ to~$w$.
It is easy to see that the approval line is not moved in any of the
other voters according to Rule~\ref{rul:preference-rewrite-rule};
in particular, the approval line is not moved in any of the voters
from the first and second group, since $B' \cap S_i \neq \emptyset$
for each~$i$, $1 \leq i \leq n$.
So $w$ is the unique SP-AV winner of election $(B' \cup
\{w\},V)$.  Since $B' \cup \{w\} = C - (B-B')$, it follows from $\|B\|
= m$ and $\|B'\| = k$ that deleting $m-k$ candidates from $C$ makes
$w$ the unique SP-AV winner.

From right to left: Let $D \subseteq B$ be any set such that $\|D\|
\leq m-k$ and $w$ is the unique SP-AV winner of election $(C-D,V)$.
Let $B' = (C-D) - \{w\}$.  Note that $B' \seq B$ and that we have the
following scores in $(B' \cup \{w\},V)$:
\begin{eqnarray*}
\scoresub{(B' \cup \{w\},V)}{w} & = & 
2(n - \ell) (k + 1) + 2 (m - \|B'\|) + 3, \\
\scoresub{(B' \cup \{w\},V)}{b_j} & \leq & 
2n (k + 1) + 2 \ell (k + 1) + 2 + 2 (m - k)
 \quad \text{ for each } b_j \in B',
\end{eqnarray*}
where $\ell$ is the number of sets $S_i \in \mathcal{S}$ that are not
hit by~$B'$, i.e., $B' \cap S_i = \emptyset$.  Recall that for
each~$i$, $1 \leq i \leq n$, all of the $2(k+1)$ voters of the form
$\begin{array}{c@{\ \ }c@{\ \ }c@{\ \ }c} S_i & | & (B-S_i) & w
\end{array}$ in the first voter group have ranked the candidates in
the same order.  Thus, for each $i$, $1 \leq i \leq n$, whenever $B'
\cap S_i = \emptyset$ one and the same candidate in $B'$ benefits from
moving the approval line according to
Rule~\ref{rul:preference-rewrite-rule},
namely the candidate occurring first in our
fixed ordering of $B'$.  Call this candidate~$b$ and note that
\begin{eqnarray*}
\scoresub{(B' \cup \{w\},V)}{b} & = & 
2n (k + 1) + 2 \ell (k + 1) + 2 + 2 (m - k).
\end{eqnarray*}
Since $w$ is the unique SP-AV winner of $(B' \cup \{w\},V)$, $w$ has
more approvals than any candidate in $B'$ and in particular more
than~$b$.  Thus, we have
\begin{eqnarray*}
\lefteqn{\scoresub{(B' \cup \{w\},V)}{w} -
\scoresub{(B' \cup \{w\},V)}{b}} \\
 & = & 2(n - \ell) (k + 1) + 2(m - \|B'\|) + 3 
- 2n (k + 1) - 2 \ell (k + 1) - 2 - 2(m - k) \\
 & = & 1 + 2 (k - \|B'\|) - 4 \ell (k + 1) ~ > ~ 0.
\end{eqnarray*}
Solving this inequality for $\ell$, we obtain
\[
0 \leq \ell < \frac{1 + 2 (k - \|B'\|)}{4 (k + 1)}
<  \frac{4 + 4k}{4 (k + 1)} = 1.
\]
Thus $\ell = 0$.  It follows that $1 + 2 (k - \| B' \|) > 0$, which
implies $\|B'\| \leq k$.  Thus, $B'$ is a hitting set of size at most
$k$.~\end{proofs}

\subsection{Voter Control}
\label{sec:results:voter-control}

Turning now to control by adding and by deleting voters, it is known
from~\cite{hem-hem-rot:j:destructive-control} that approval voting is
resistant to constructive control and is vulnerable to destructive
control (see Table~\ref{tab:summary-of-results}).\footnote{Meir et
al.~\cite{mei-pro-ros-zoh:j:multiwinner}
proved in
their interesting ``multi-winner'' model (which generalizes Bartholdi,
Tovey, and Trick's model~\cite{bar-tov-tri:j:control} by adding a
utility function and some other parameters) that approval voting is
resistant to constructive control by adding voters.  According to
Footnote~13 of \cite{hem-hem-rot:j:destructive-control}, this
resistance result immediately follows from the corresponding
resistance result in
\cite{hem-hem-rot:c:destructive-control,hem-hem-rot:j:destructive-control},
essentially due to the fact that lower bounds in more flexible models
are inherited from more restrictive models.
\label{foo:multi-winner}}
Their proofs can be modified so as to also apply to sincere-strategy
preference-based approval voting.  We here provide only proof sketches;
more details of the proofs are
provided in the technical report version~\cite{erd-now-rot:t:sp-av}.

\begin{theorem}
\label{thm:adding-deleting-voters}
SP-AV is resistant to constructive control by adding voters and by
deleting voters and is vulnerable to destructive control by adding
voters and by deleting voters.
\end{theorem}

\sproofsketchof{Theorem~\ref{thm:adding-deleting-voters}}
Susceptibility holds by Lemma~\ref{lem:susceptible-voter-control} in
all cases.  To prove resistance to constructive control by adding
voters (respectively, by deleting voters), the construction of
\cite[Thm.~4.43]{hem-hem-rot:j:destructive-control} (respectively, of
\cite[Thm.~4.44]{hem-hem-rot:j:destructive-control}) works, modified
only by specifying voter preferences consistently with the voters'
approval strategies (and, in the deleting-voters case, by adding a
dummy candidate who is disapproved of and ranked last by every voter in the
construction to ensure an admissible AV strategy profile).  These
constructions provide polynomial-time reductions from the
$\np$-complete problem Exact Cover by Three-Sets (denoted by X3C; see,
e.g., Garey and Johnson~\cite{gar-joh:b:int}), which is defined as
follows: \begin{desctight}

\item[Name] Exact Cover by Three-Sets (X3C).

\item[Instance] A set $B = \{b_1, b_2, \ldots , b_{3m}\}$,
$m>1$,\footnote{Our assumption
that $m>1$ is not explicitly specified in Garey and
Johnson~\cite{gar-joh:b:int}.  However, it is clear that requiring
$m>1$ does not change the complexity of the problem.} and
a collection $\mathcal{S} = \{S_1, S_2, \ldots , S_n\}$ of subsets
$S_i \seq B$ with $\|S_i\| = 3$ for each~$i$.

\item[Question] Does $\mathcal{S}$ have an exact cover for~$B$, i.e.,
is there a subcollection $\mathcal{S}' \seq \mathcal{S}$ such that
every element of $B$ occurs in exactly one set in~$\mathcal{S}'$?
\end{desctight}

The polynomial-time algorithms showing that approval voting is
vulnerable to destructive control by adding voters and by deleting
voters~\cite[Thm.~4.24]{hem-hem-rot:j:destructive-control} can be
straightforwardly adapted to also work for sincere-strategy
preference-based approval voting, since no approval lines are moved 
according to Rule~\ref{rul:preference-rewrite-rule} in
these control
scenarios.~\qed

We now prove that, just like plurality,
sincere-strategy preference-based approval voting is
resistant to constructive and destructive control by partition of
voters in model~{TP}.  In fact, the proof presented
in~\cite{hem-hem-rot:j:destructive-control} for plurality in these two
cases also works for {SP-AV} with minor modifications.
In contrast, approval voting is vulnerable
to the destructive variant of this control
type~\cite{hem-hem-rot:j:destructive-control}.

\jfootnote{Add citation \cite{hem-hem-rot:j:destructive-control} for
Theorem~\ref{thm:resistance-voter-partition-TP} also in MFCS-Version.}

\begin{theorem}
\label{thm:resistance-voter-partition-TP}
SP-AV is resistant to constructive and destructive control by
partition of voters in model~{TP}.
\end{theorem}

\sproofsketchof{Theorem~\ref{thm:resistance-voter-partition-TP}} The
proof is again based on
Construction~\ref{con:resistance-general-candidate-control}, but the
reduction is now from Restricted Hitting Set, which is defined just as
Hitting Set (see Section~\ref{sec:results:candidate-control}) except
that $n(k+1) + 1 \leq m-k$ is required in addition.  Restricted
Hitting Set is also
$\np$-complete~\cite{hem-hem-rot:j:destructive-control}.  Now, the key
observation is the following proposition, which can be proven as
in~\cite{hem-hem-rot:j:destructive-control}.

\begin{proposition}[Hemaspaandra et 
al.~\cite{hem-hem-rot:j:destructive-control}]
\label{cla:resistance-voter-control-TP}
Let $(B,\mathcal{S},k)$ be a given Restricted Hitting Set instance, where $B =
\{b_1, b_2, \ldots , b_m\}$ is a set, $\mathcal{S} = \{S_1, S_2,
\ldots , S_n\}$ is a nonempty collection of subsets $S_i \seq B$, and $k \leq
m$ is a positive integer such that $n(k+1) + 1 \leq m-k$.  If $(C,V)$
is the election resulting from $(B,\mathcal{S},k)$ via
Construction~\ref{con:resistance-general-candidate-control}, then the
following three statements are equivalent:
\begin{enumerate}
\item $\mathcal{S}$ has a hitting set of size less than or equal
  to~$k$.
\item $V$ can be partitioned such that $w$ is the unique SP-AV winner
  in model~{TP}.
\item $V$ can be partitioned such that $c$ is not the unique SP-AV
  winner in model~{TP}.
\end{enumerate}
\end{proposition}
The theorem now follows immediately from
Proposition~\ref{cla:resistance-voter-control-TP}.~\eproofof{Theorem~\ref{thm:resistance-voter-partition-TP}}

Finally, we turn to control by partition of voters in model~{TE}.  For
this control type, Hemaspaandra et
al.~\cite{hem-hem-rot:j:destructive-control} proved approval voting
resistant in the constructive case and vulnerable in the destructive
case.  We have the same results for sincere-strategy preference-based
approval voting.  Our resistance proof in the constructive case
(see the proof of Theorem~\ref{thm:resistance-voter-partition-TE}) is
similar to the corresponding proof of resistance
in~\cite{hem-hem-rot:j:destructive-control}.  However, while our
polynomial-time algorithm showing vulnerability for SP-AV in the
destructive case (see the proof of 
Theorem~\ref{thm:vulnerability-voter-partition-TE})
is based on the corresponding polynomial-time
algorithm for approval voting
in~\cite{hem-hem-rot:j:destructive-control}, it extends their
algorithm in a nontrivial way.

\begin{theorem}
\label{thm:resistance-voter-partition-TE}
SP-AV is resistant to constructive control by
partition of voters in model~{TE}.
\end{theorem}

\begin{proofs}
Susceptibility holds by Lemma~\ref{lem:susceptible-voter-control}.
The proof of resistance is based on the construction of
\cite[Thm.~4.46]{hem-hem-rot:j:destructive-control} with only minor
changes.  Let an X3C instance $(B,\mathcal{S})$ be given, where $B =
\{b_1, b_2, \ldots , b_{3m}\}$, $m>1$, is a set and $\mathcal{S} =
\{S_1, S_2, \ldots , S_n\}$ is a collection of subsets $S_i\seq B$
with $\|S_i\| = 3$ for each~$i$, $1\leq i \leq n$.
Without loss of generality, we may
assume that $n \geq m$.  Define the value $\ell_j = \| \{S_i \in
\mathcal{S} \condition b_j\in S_i \}\|$ for each~$j$, $1\leq j \leq
3m$.

Define the election $(C,V)$, where $C = B \cup \{w,x,y\} \cup Z$ is
the candidate set with the distinguished candidate~$w$, $Z = \{z_1,
z_2, \ldots ,z_n\}$, and where $V$ is defined to consist of the
following $4n+m$ voters:
\begin{enumerate}
\item For each~$i$, $1 \leq i \leq n$, there is one voter of
  the form:
$\begin{array}{c@{\ \ }c@{\ \ }c@{\ \ }c@{\ \ }c}
y & S_i & | & w & ((B-S_i)\cup \{ x \} \cup Z) .
\end{array}$
\item For each~$i$, $1 \leq i \leq n$, there is one voter of
  the form:
$\begin{array}{c@{\ \ }c@{\ \ }c@{\ \ }c@{\ \ }c}
y & z_i & | & w & (B \cup \{ x \} \cup (Z- \{z_i \} )) .
\end{array}$
\item For each~$i$, $1 \leq i \leq n$, there is one voter of the form:
$\begin{array}{c@{\ \ }c@{\ \ }c@{\ \ } c@{\ \ }c@{\ \ }c@{\ \ }c@{\ \
}c} w & (Z- \{ z_i \} ) & B_i & | & x & y & z_i & (B-B_i) ,
\end{array}$
where $B_i = \{ b_j \in B \condition i\leq n-\ell_j \}$.

\item There are $n+m$ voters of the form:
$\begin{array}{c@{\ \ }c@{\ \ }c@{\ \ }c}
x & | & y & (B\cup \{ w \} \cup Z).
\end{array}$
\end{enumerate}

Note that $\scoresub{(C,V)}{b_j} = n$ for each $b_j \in B$.
Since the above construction is only slightly modified from the proof
of \cite[Thm.~4.46]{hem-hem-rot:j:destructive-control}, so as to
formally conform with the SP-AV voter representation, the same
argument as in that proof shows that $\mathcal{S}$ has an exact cover
for $B$ if and only if $w$ can be made the unique SP-AV winner by
partition of voters in model~{TE}.  Note that, in the present control
scenario, approval voting and SP-AV can differ only in the run-off,
but the construction ensures that they don't differ there.

From left to right, if $\mathcal{S}$ has an exact cover for $B$ then
partition the set of voters as follows: $V_1$ consists of the $m$
voters of the form
$\begin{array}{@{}c@{\ \ }c@{\ \ }c@{\ \ }c@{\ \ }c@{}}
 y & S_i & | & w & ((B-S_i)\cup \{ x \} \cup Z)
\end{array}$
that correspond to the sets in the exact cover, of the $n+m$ voters
who approve of only~$x$, and of the $n$ voters who approve of $y$
and~$z_i$, $1 \leq i \leq n$. Let $V_2 = V - V_1$.  It follows that
$w$ is the unique SP-AV winner of both subelection $(C,V_2)$ and the
run-off, simply because no candidate proceeds to the run-off from the
other subelection, $(C,V_1)$, in which $x$ and $y$ tie for winner with
a score of $n+m$ each.

From right to left, suppose $w$ can be made the unique SP-AV winner by
partition of voters in model~{TE}.  Let $(V_1,V_2)$ be a partition of
$V$ such that $w$ is the unique SP-AV winner of the run-off.
According to model TE, $w$ must also be the unique SP-AV winner of one
subelection, say of $(C,V_1)$.
Note that each voter of the form 
$\begin{array}{@{}c@{\ \ }c@{\ \ }c@{\ \ }c@{\ \ }c@{}}
y & z_i & | & w & (B \cup \{ x \} \cup (Z- \{z_i \} ))
\end{array}$
has to be in $V_2$ (otherwise, we would have $\scoresub{(C,V_1)}{w} = 
\scoresub{(C,V_1)}{z_i} $ for at least one~$i$, and so $w$ would not be
the unique SP-AV winner of $(C,V_1)$ anymore).
However, if there were more than $m$
voters of the form $\begin{array}{@{}c@{\ \ }c@{\ \ }c@{\ \ }c@{\ \
}c@{}} y & S_i & | & w & ((B-S_i)\cup \{ x \} \cup Z)
\end{array}$
in $V_2$ then $\scoresub{(C,V_2)}{y}>n+m $, and so $y$ would be the
unique SP-AV winner of the other subelection, $(C,V_2)$.  But then,
also in the SP-AV model, $y$ would win the run-off against $w$ because
$\scoresub{(\{w,y\},V)}{y} = 3n+m > n = \scoresub{(\{w,y\},V)}{w}$,
which contradicts the assumption that $w$ has been made the unique
SP-AV winner by the partition $(V_1,V_2)$.  Hence, there are at most
$m$ voters of the form $\begin{array}{@{}c@{\ \ }c@{\ \ }c@{\ \ }c@{\
\ }c@{}} y & S_i & | & w & ((B-S_i)\cup \{ x \} \cup Z)
\end{array}$
in~$V_2$, and these $m$ voters correspond to an exact cover
of~$B$, since otherwise there would be at least one $b_j \in B$ with
$\scoresub{(C,V_1)}{b_j} = n = \scoresub{(C,V_1)}{w}$.~\end{proofs}

\begin{theorem}
\label{thm:vulnerability-voter-partition-TE}
SP-AV is vulnerable to destructive control by
partition of voters in model~{TE}.
\end{theorem}

\begin{proofs}
Susceptibility holds by Lemma~\ref{lem:susceptible-voter-control}.  To
prove vulnerability, we describe a polynomial-time algorithm showing
that (and how) the chair can exert destructive control by partition of
voters in model TE for sincere-strategy preference-based approval
voting.  Our algorithm extends the polynomial-time algorithm designed
by Hemaspaandra et al.~\cite{hem-hem-rot:j:destructive-control} to
prove approval voting vulnerable to this type of control.
Specifically, our algorithm adds Loop~2 below to their algorithm, and
we will explain below why it is necessary to add this second loop.

Let $(C,V)$ be an
election, and for each voter $v \in V$, let $S_v \seq C$ denote $v$'s
AV strategy. In each iteration of Loop~1 in the algorithm below, we
will consider three candidates, $a$, $b$, and~$c$.  Define the
following five numbers:\footnote{This notation is adopted
from~\cite{hem-hem-rot:j:destructive-control} and adjusted here to the
SP-AV system.  $W_c$ is the number of votes in which $c$ \emph{w}ins
one approval against both $a$ and~$b$,
$L_c$ is the number of votes in which $c$ \emph{l}oses one approval
against both $a$ and~$b$, and $D_a$, $D_b$, and $D_{ac}$
are the numbers of votes in which the candidate(s) in the subscript
gain one approval against the candidate(s) not in the subscript (thus
decreasing their \emph{d}eficit).}
\begin{eqnarray*}
W_c   = \|\{ v \in V \condition
a \not\in S_v,\ b \not\in S_v,\ c \in S_v\}\|, & & 
L_c   = \|\{ v \in V \condition
a \in S_v,\ b \in S_v,\ c \not\in S_v\}\|, \\
D_a   = \|\{ v \in V \condition
a \in S_v,\ b \not\in S_v,\ c \not\in S_v\}\|, & & 
D_b   = \|\{ v \in V \condition
a \not\in S_v,\ b \in S_v,\ c \not\in S_v\}\|, \mbox{ and}\\
D_{ac}   = \|\{ v \in V \condition
a \in S_v,\ b \not\in S_v,\ c \in S_v\}\|. & & 
\end{eqnarray*}

In addition, we introduce the following notation.  Given an election
$(C,V)$ and two distinct candidates $x, y \in C$, let $\diff(x,y)$
denote the number of voters in $V$ who prefer $x$ to $y$ minus the
number of voters in $V$ who prefer $y$ to~$x$.  Define $B_c$ to be the
set of candidates $y \neq c$ in $C$ such that $\diff(y,c) \geq 0$.

The input to our algorithm is an election $(C,V)$, where each voter $v
\in V$ has a sincere AV strategy $S_v$
(otherwise, the input is considered malformed and outright
rejected), and a distinguished candidate $c \in C$.  On this input,
our algorithm works as follows.

\begin{enumerate}
\item {\bf Checking the trivial cases:} can be done as in the case of
approval voting, see the proof of
\cite[Thm.~4.21]{hem-hem-rot:j:destructive-control}.  In particular,
if $C = \{c\}$ then output ``control impossible'' and halt, since $c$
cannot help but win.  If $C$ contains more candidates than only $c$
but $c$ already is not the unique SP-AV winner of $(C,V)$ then output
the (successful) partition $(V,\emptyset)$ and halt.  Otherwise, if
$\|C\| = 2$ then output ``control impossible'' and halt, as $c$ is the
unique SP-AV winner of $(C,V)$ in the current case and so, however the
voters are partitioned, $c$ must win---against the one rivalling
candidate---at least one subelection and also the run-off.

\item {\bf Loop~1:} For each $a, b \in C$ such that $\|\{a, b, c\}\| =
3$, check whether $V$ can be partitioned into $V_1$ and $V_2$ such
that $\scoresub{(C,V_1)}{a} \geq \scoresub{(C,V_1)}{c}$ and
$\scoresub{(C,V_2)}{b} \geq \scoresub{(C,V_2)}{c}$.  As shown in the
proof of \cite[Thm.~4.21]{hem-hem-rot:j:destructive-control}, this is
equivalent to checking
\begin{eqnarray}
\label{eq:test-loop1}
W_c - L_c \leq D_a + D_b.
\end{eqnarray}
If (\ref{eq:test-loop1}) fails, this $a$ and $b$ cannot prevent $c$
from being the unique winner of at least one subelection and thus also
of the run-off, so we move on to test the next $a$ and $b$ in this
loop.  If (\ref{eq:test-loop1}) holds, however, output the partition
$(V_1, V_2)$ and halt, where $V_1$ consists of the voters contributing
to $D_a$, of the voters contributing to $D_{ac}$, and of $\min(W_c,
D_a)$ voters contributing to $W_c$, and where $V_2 = V - V_1$.

\item {\bf Loop~2:} For each $d \in B_c$, partition $V$ as follows.
Let $V_1$ consist of all voters in $V$ who approve of~$d$, and let
$V_2 = V - V_1$.  If $d$ is the unique winner of $(C, V_1)$, then
output $(V_1, V_2)$ as a successful partition and halt.  Otherwise, go
to the next $d \in B_c$.

\item {\bf Termination:} If in no iteration of either Loop~1 or Loop~2
a successful partition of $V$ was found, then output ``control
impossible'' and halt.
\end{enumerate}

Let us give a short explanation of why Loop~2 is needed for {SP-AV} by
stressing the difference with approval voting. As shown in the proof
of \cite[Thm.~4.21]{hem-hem-rot:j:destructive-control}, if none of the
trivial cases applied, then condition~(\ref{eq:test-loop1}) holds for
some $a, b \in C$ with $\|\{a, b, c\}\| = 3$ if and only if
destructive control by partition of voters in model~TE is possible for
approval voting.  Thus, for approval voting, if Loop~1 was not
successful for any such $a$ and $b$, we may immediately jump to the
termination stage, where the algorithm outputs ``control impossible''
and halts.  In contrast, if none of the trivial cases applied, then
the existence of candidates $a$ and $b$ with $\|\{a, b, c\}\| = 3$ who
satisfy (\ref{eq:test-loop1}) is \emph{not} equivalent to destructive
control by partition of voters in model~TE being possible for SP-AV:
It is a sufficient, yet not a necessary condition.  The reason is that
even if there are no candidates $a$ and $b$ who can prevent $c$ from
winning one subelection (in some partition of voters) and from
proceeding to the run-off, it might still be possible that $c$ loses
or ties the run-off due to moving the approval line according to
Rule~\ref{rul:preference-rewrite-rule}.

Indeed, if Loop~1 was not successful, $c$ will lose or tie the
run-off exactly if there exists a candidate $d \neq c$ such that
$\diff(d,c) \geq 0$ and $d$ can win one subelection (for some
partition of voters).  This is precisely what is being checked in
Loop~2.  Indeed, note that the partition $(V_1, V_2)$ chosen in Loop~2
for $d \in B_c$ is the best possible partition for $d$ in the
following sense: If $d$ is not a unique SP-AV winner of subelection
$(C,V_1)$ then, for each $W \seq V$, $d$ is not a unique SP-AV
winner of subelection $(C,W)$.  To see this, simply note that if $d$
is not a unique SP-AV winner of $(C,V_1)$, then there is some
candidate $x$ with $\scoresub{(C,V_1)}{x} = \scoresub{(C,V_1)}{d} =
\|V_1\|$, which by our choice of $V_1$ implies $\scoresub{(C,W)}{x}
\geq \scoresub{(C,W)}{d}$ for each subset $W \seq V$.~\end{proofs}

\section{Conclusions and Open Questions}
\label{sec:conclusions}

We have shown that Brams and Sanver's sincere-strategy
preference-based approval voting
system~\cite{bra-san:j:critical-strategies-under-approval}, when
adjusted so as to \emph{coerce} admissibility (rather than excluding 
inadmissible votes \emph{a priori}), combines the resistances of
approval and plurality voting to procedural control: SP-AV is
resistant to 19 of the 22 previously studied types of control.  On the
one hand, like Copeland voting
\cite{fal-hem-hem-rot:j-To-Appear-With-TR-Ptr:llull-copeland-full-techreport,fal-hem-hem-rot:c:copeland-fully-resists-constructive-control},
SP-AV is fully resistant to constructive control, yet unlike Copeland
it additionally is broadly resistant to destructive control.  On the
other hand, like
plurality~\cite{fal-hem-hem-rot:j-To-Appear-With-TR-Ptr:llull-copeland-full-techreport,bar-tov-tri:j:control,hem-hem-rot:j:destructive-control,fal-hem-hem-rot:c:llull},
SP-AV is fully resistant to candidate control, yet unlike plurality it
additionally is broadly resistant to voter control.  In conclusion, for these
$22$ types of control, SP-AV has more resistances
and fewer
vulnerabilities to control than is currently known for any other
natural voting system with a polynomial-time winner problem (see
Table~\ref{tab:number-of-resistances}).  However, when comparing
approval voting and SP-AV, it should also be noted that the former
is even immune to nine of these $22$ control types, whereas the
latter has no immunities at all.  Since immunity may be seen as a perfect
protection against control and resistance provides protection to
control only in a computational sense, one should carefully evaluate
the pros and cons of both systems.  The result of such an evaluation
will certainly depend on which particular types of control one wishes
to be protected against.

As an interesting task for future research, we propose to expand the
study of SP-AV with respect to other computational properties than its
behavior regarding procedural control (see, e.g.,
\cite{fal-hem-hem-rot:b:richer,bau-erd-hem-hem-rot:btoappear-m:computational-apects-of-approval-voting}),
and to investigate also its social choice properties in more detail.
In addition, we propose as an interesting and extremely
ambitious task for future work the study of SP-AV (and other voting
systems as well) beyond the worst-case---as we have done here---and
towards an appropriate typical-case complexity model; see, e.g.,
\cite{mcc-pri-sli:j:dodgson,pro-ros:j:juntas,con-san:c:nonexistence,hem-hom:jtoappearWithPtrFull:dodgson-greedy,erd-hem-rot-spa:c:lobbying}
for interesting results and discussion in this direction.

\medskip
\noindent
{\bf Acknowledgments:} We are grateful to Edith and Lane A. Hemaspaandra 
for helpful comments and interesting discussions that are reflected in parts of
Section~\ref{sec:discussion-spav}.
We thank the anonymous MLQ, MFCS-08, and COMSOC-08
referees for their helpful comments on preliminary versions of this
paper.

{\small 
\bibliographystyle{alpha}

\providecommand{\WileyBibTextsc}{}
\let\textsc\WileyBibTextsc
\providecommand{\othercit}{}
\providecommand{\jr}[1]{#1}
\providecommand{\etal}{~et~al.}

\end{document}